\documentclass{aa}  

\usepackage{graphicx}
\usepackage{txfonts}
\usepackage{xcolor}
\usepackage{mathrsfs}  
\usepackage{paralist}
\usepackage{wasysym}
\usepackage{listings}
\usepackage{stackengine}
\renewcommand\dot[1]{\stackMath\stackengine{1pt}{#1}{\mbox{\large\bfseries .}}{O}{c}{F}{T}{S}}

\let\oldsqrt\sqrt
\renewcommand\sqrt[1]{\!{\textstyle\oldsqrt{#1}}\,}
\def\sqrt#1{\!\oldsqrt{#1}\,}
\usepackage{natbib,twoopt}
\usepackage[breaklinks=true]{hyperref} 
\bibpunct{(}{)}{;}{a}{}{,}             
\makeatletter
  \newcommandtwoopt{\citeads}[3][][]{\href{http://adsabs.harvard.edu/abs/#3}%
    {\def\hyper@linkstart##1##2{}%
     \let\hyper@linkend\@empty\citealp[#1][#2]{#3}}}
  \newcommandtwoopt{\citepads}[3][][]{\href{http://adsabs.harvard.edu/abs/#3}%
    {\def\hyper@linkstart##1##2{}%
     \let\hyper@linkend\@empty\citep[#1][#2]{#3}}}
  \newcommandtwoopt{\citetads}[3][][]{\href{http://adsabs.harvard.edu/abs/#3}%
    {\def\hyper@linkstart##1##2{}%
     \let\hyper@linkend\@empty\citet[#1][#2]{#3}}}
  \newcommandtwoopt{\citeyearads}[3][][]%
    {\href{http://adsabs.harvard.edu/abs/#3}
    {\def\hyper@linkstart##1##2{}%
     \let\hyper@linkend\@empty\citeyear[#1][#2]{#3}}}
\makeatother

\begin{document}

\title{Collision detection for $N$-body Kepler systems}
\author{P.M.\ Visser}
\institute{
Delft Institute of Applied Mathematics, Delft University of Technology,
Mekelweg 4, 2628 CD Delft, The Netherlands \\
\email{p.m.visser@tudelft.nl}
}
\date{Received April 11, 2022\ / Accepted October 29, 2022}
\abstract{In a Keplerian system, a large number of bodies orbit a central mass. Accretion disks, protoplanetary disks, asteroid belts, and planetary rings are examples. Simulations of these systems require algorithms that are computationally efficient. The inclusion of collisions in the simulations is challenging but important.
}
{We intend to calculate the time of collision of two astronomical bodies in intersecting Kepler orbits  as a function of the orbital
elements. The aim is to use the solution in an analytic propagator ($N$-body simulation) that jumps from one collision event to the next.
}
{We outline an algorithm that maintains a list of possible collision pairs ordered chronologically. At each step (the soonest event on the list), only the particles created in the collision can cause new collision possibilities. We estimate the collision rate, the length of the list, and the average change in this length at an event, and study the efficiency of the  method used.
}
{We find that the collision-time problem is equivalent to finding the grid point between two parallel lines that is closest to the origin. The solution is based on the continued fraction of the ratio of orbital periods.
}
{Due to the large jumps in time, the algorithm can beat tree codes (octree and $k$-d tree codes can efficiently detect collisions) for specific systems such as the Solar System with $N<10^8$. However, the gravitational interactions between particles can only be treated as gravitational scattering or as a secular perturbation, at the cost of reducing the time-step or at the cost of accuracy. While simulations of this size with high-fidelity propagators can already span vast timescales, the high efficiency of the collision detection allows many runs from one initial state or a large sample set, so that one can study statistics.
}

\maketitle


\section{Introduction}
Simulations of the mechanical motion of many bodies are generally computationally expensive. Consider the $N$-body problem. Here, each of the $N$ particles moves under the influence of the gravitation of all other particles and each particle can collide with any other particle. Therefore, there are $N^2$ interactions to account for.
Various methods have been invented to speed up the simulation or increase the particle count:
\begin{inparaenum}[(i)]
\item
direct $N$-body simulations with dynamic time-steps \citep[see][for an overview]{Dehnen2011};
\item
the octree code for collision detection \citep{Bentley1975,Meagher1980};  
\item
the \emph{Barnes-Hut algorithm} \citep{Hut1986,Barnes1990,Hamada2009,Burtscher2011} for mutual gravity, where nearby particles are grouped so that their effect on a distant particle can be combined, which requires $O(N\log N)$ computational steps;
\item
the fast multi-pole \emph{Greengard and Rokhlin method} (FMM), where higher order moments of the particle groups are included \citep{Rokhlin1985,Greengard1990};
\item
parallelization of these methods \citep{Warren1993};
\item
particle mesh methods, where the $N$ force vectors are calculated using the Newton potential and the Poisson Equation~for the potential is solved numerically with fast Fourier transforms \citep{Bodenheimer2007};
\item
the finite-elements method (FEM); and finally, 
\item
for a Keplerian system (with a large central mass), where the particles move in slowly precessing Kepler ellipses described by the \emph{Laplace-Lagrange equations} for the orbital elements \citep{Murray2009}; because the Kepler ellipses change slowly over time, the time-step in these numerical integration propagators can be many orbital periods.
\end{inparaenum}

In astronomy, collision detection is the problem of finding the precise moment at which asteroids, planets, or satellites collide. Here the difficulty in predicting collisions, or calculating the collision probability, stems from the fact that the objects are very small compared to the size of their orbits.
In numerical simulations the number of nearby particles that need to be considered is a function of the step size. Because the particles travel during each time-step,
the volume of space around the particle that needs to be probed for collision partners has a radius of the time-step times velocity. In order to limit the number
of collision partners, this volume needs to remain small. Therefore, the step size decreases as $O(N^{-1/3})$ and the total number of steps for a fixed simulation time grows as $O(N^{1/3})$. Efficient codes, such as octree codes \citep{Meagher1982} or spatial hashing codes, have an algorithmic efficiency of $O(N\log N)$ per time-step. If these are used in collision detection, the number of steps grows as $O(N^{4/3}\log N)$. This makes the problem of collision detection in astronomy even more challenging than pure gravitational evolution without collisions \citep[see][for a comparison between codes with and without collision detection]{Dehnen2011}.

In this paper, we apply collision detection to Keplerian systems, such as astrophysical disks, where all particles feel a dominant gravity force from one heavy central mass. Each particle is in a Kepler orbit given by parameters $a$, $\epsilon$, $\varpi$, $I$, $\ascnode$, and $\nu$ (see Table~\ref{table1} for the symbols). However, the advantage of implementing collision detection in an analytic propagator is that the algorithm can be very efficient. In Sect.~\ref{Sec2}, we analyze the timescales, evaluate the numerical efficiency, and compare it with algorithms based on tree code. As there is no numerical integration of the orbits, many physical effects are neglected (see Sect.~\ref{Sec2.3}).

\begin{figure}
\centering
\includegraphics{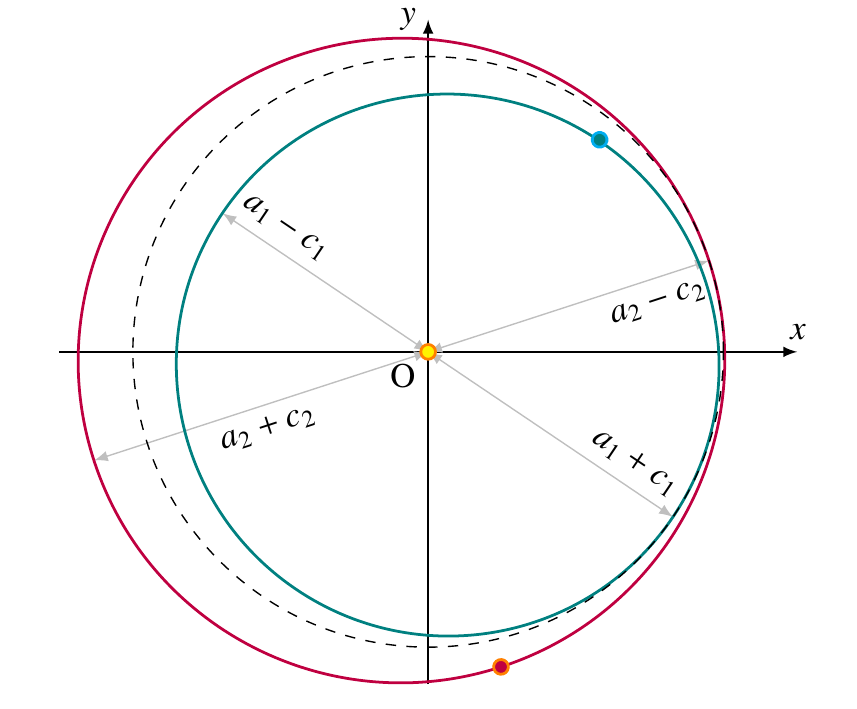}
\caption{Two orbits separated by a sphere (\emph{dashed}). Orbit~1 (\emph{blue}) has apoapsis $a_1+c_1$ and orbit~2 (\emph{purple}) has periapsis $a_2-c_2$. Filtering out such collision-avoiding apoapsis--periapsis pairs is an efficient sweep and prune method.}
\label{Figure1}
\end{figure}

\begin{figure}[t]
\centering
\includegraphics{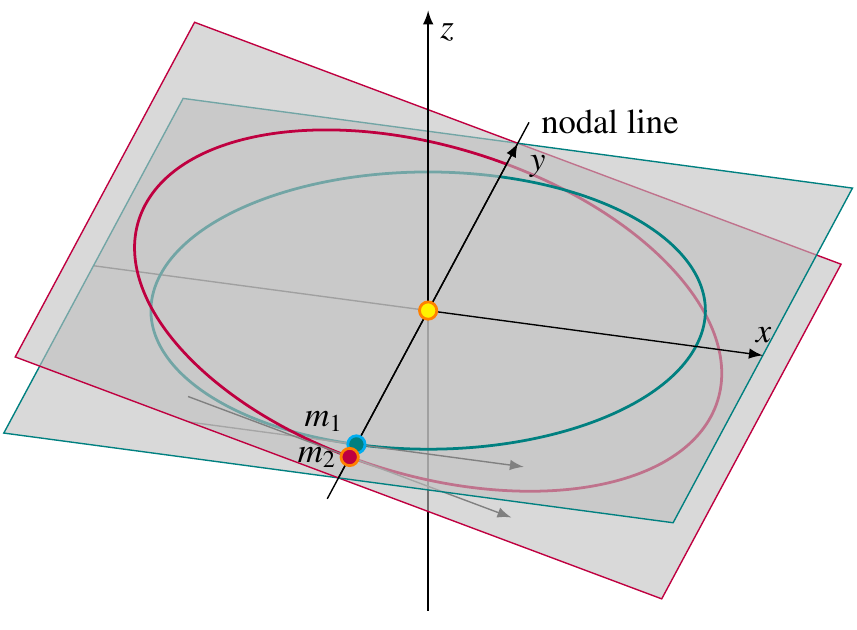}
\caption{Orbits of two planetoids $m_1$, $m_2$ in their orbital planes. Because the bodies are much smaller than the orbits, $s_j\ll a_j$, collisions happen near the mutual nodal line of intersection of the orbital planes, even for small inclinations. The tangent vectors (\emph{gray}) indicate the linear approximation that may be used to find the collision point.}
\label{Figure2}
\end{figure}

We describe (in Sect.~\ref{Sec3}) the algorithm for an $N$-body code with collision detection  in detail. Initially, it compares particles
sorted by radial distance using equations from the seminal paper by \citet{Opik1951}; see Fig.~\ref{Figure1}. This is effectively an implementation of the apoapsis/periapsis filter of \citet{Hoots1984} and an example of a sweep and prune method. The algorithm then uses analytic evaluation of the points of collision \citep[near the nodal line in Fig.~\ref{Figure2}, following][as explained in Sect.~\ref{sec4}]{Hoots1984,Manley1998}, of the earliest crossing time (Sect.~\ref{Sec5}), and of the time of collision (derived in Sect.~\ref{Sec6}).
The algorithm keeps track of pairs of particles that are on a collision course, from the earliest to the latest moment of collision.
Each step of the simulation involves only the calculation of the next collision, and updating the list. In the method, time-steps increase with decreasing $s$, which allows long simulation times. Indeed, for the limiting case $s\longrightarrow 0$, the algorithm stops after initialization, as it finds that there are no collisions. In contrast, collision detection using a numerical integration propagator always requires a nearest-neighbor search for every particle. The time spent on this search is independent of $s$.

To our knowledge, the idea of bookkeeping a list of future possible collisions has not been studied elsewhere. The algorithm relies on a novel method to quickly find the exact collision times. We derived new expressions, Eqs.~(\ref{cos}) and (\ref{sin}), for the difference in the eccentric anomaly between two given points on an orbit that are also accurate at small eccentricities, when the eccentric anomalies themselves are ill defined. These formulas were needed to calculate the time for a particle to get to the collision point.

\begin{table}[t]
\caption{List of symbols and notation.}
\label{table1}      
\centering
\begin{tabular}{l|l}
\hline
\hline
symbol & quantity \\
\hline
$t$ & time \\
$\mathrm dt$ & time-step, small time interval \\
$x$, $y$, $z$ & Cartesian coordinates \\
$V$ & volume \\
$h=\Delta a$ & thickness of spherical shell \\
\hline
$\boldsymbol r=\begin{pmatrix} x \\ y \\ z \end{pmatrix}$ & position vector \\
$r$ & radial distance \\
$\boldsymbol v=\dot{\boldsymbol r}$ & velocity vector \\
$v$ & speed \\ 
$\boldsymbol d=\boldsymbol r_2-\boldsymbol r_1$ & difference position \\
$d$ & distance \\
$\boldsymbol u=\boldsymbol v_2-\boldsymbol v_1$ & difference velocity \\
$u$ & relative speed  \\
$\boldsymbol w=\boldsymbol v_1\times\boldsymbol v_2$ & \\
\hline
$G$ & Newton's constant \\
$M$ & mass of central body \\
$S$ & radius of central body \\
\hline
$m$ & particle mass \\
$s$ & particle radius \\
$t^0$ & particle creation time \\
$T$ & orbit time \\
$\omega=2\pi/T$ & mean motion \\
$T_\text{coll}$ & time between collisions \\
$T_\text{scat}$ & time between close encounters \\
$T_\text{prec}$ & precession period/time scale \\
$T_\text{sim}$ & total simulated time \\
\hline
$a$, $b$ & semimajor-, semiminor axis \\
$c=a\epsilon$ & semi-focal separation \\
$\ell$ & semi-latus rectum \\
$\boldsymbol r^0$ & particle creation point \\
$\boldsymbol S$ & spin angular momentum \\
$\boldsymbol L$ & orbital angular momentum \\
$\boldsymbol K =\boldsymbol L_1\times \boldsymbol L_2$ & direction  of nodal line \\
$\boldsymbol\epsilon$ & eccentricity vector \\
$\varpi$ & argument of periapsis \\
$\ascnode$ & ascending node \\
$I$ & inclination \\
$\nu$ & true anomaly \\
$E$ & eccentric anomaly \\
$\mathscr{R}$ & rotation matrix \\
\hline
$N$ & number of particles \\
$\kappa$ & number of fragments \\
$i$, $j$ & particle indices \\
$k$, $l$ & rounds to collision \\
$n$ & counter \\
\hline
\hline
\end{tabular}
\tablefoot{Symbol and significance of the physical quantities used. The symbol $\varpi$, which usually represents the longitude of the periapsis is here used for the argument of periapsis, so that we can reserve the symbol $\omega$ for the angular frequency or mean motion.}
\end{table}

\section{Applications and estimates of timescales}
\label{Sec2}
The algorithm is intended for simulation of the dynamical evolution of a large planetary ring system or a debris disk around a star. In the latter, nearby passages
also occur, where one planet is inside the sphere of gravitational influence of another. The resulting gravitational scattering may be modeled with an elastic collision. These systems therefore have four different timescales: the orbit time $T$, the collision time $T_\text{coll}$, the scattering time $T_\text{scat}$, and the secular time $T_\text{prec}$.

{In order to estimate how many collisions happen per unit of time},
we first consider a thin disk with a completely randomized (homogeneous) distribution of particles in near-circular orbits. A particle with radius $s$ traces a cylindrical volume of size $(2\pi a/T)\pi s^2$ per unit of time. The disk is a cylinder of radius $a$ and height $2Ia$, meaning that the particle density is $N/2\pi Ia^3$. Accounting for the $N^2/2$ pairs, the rate of collisions is estimated to be
\[
\frac{1}{T_\text{coll}} =
\underbrace{\frac{2\pi a}{T}}_{\text{velocity}}\cdot
\underbrace{\pi (2s)^2}_{\substack{\text{effective} \\ \text{cross section}}} \cdot
\underbrace{\frac{N}{2\pi Ia^3}}_{\text{density}} \cdot
\underbrace{\frac{N}{2}}_{\text{pairs}} =
\frac{2\pi N^2s^2}{Ia^2T}
.
\]
If we want to include close encounters, we may substitute $s$ into the formula for the radius of the sphere of influence $s=(m/M)^{2/5}a$. The timescales are therefore
\[
T_\text{scatt} = \dfrac{I}{2\pi N^2} \bigg(\dfrac{M}{m}\bigg)^{4/5} T , \quad
T_\text{coll} = \dfrac{Ia^2}{2\pi N^2s^2}T , \quad
T_\text{prec} = \dfrac{4M}{Nm} T
.
\]
The formula for the precession is taken from \citet{Murray2009}. If we model the early inner Solar System by $N=10^6$ planetesimals of characteristic size $s=100\text{km}$ in a disk with $a=4\text{au}$ and $I=.1$, we have
\[
T_\text{scatt} \approx 45\text{min} , \quad
T_\text{coll} \approx 35\text{yr} , \quad
T_\text{prec} \approx 10^5\text{yr} , \quad
T \approx 10\text{yr} .
\]
Next, we consider a ring system around a planet. If we assume its radius is only a few times that of the planet and the ring particles have the same density as the planet, the collision time is comparable to the scattering time. For the Uranus ring system, we take $N=10^{13}$, $s=1\text{m}$, and $a=10^5\text{km}$, which results in
\[
T_\text{scatt} \approx T_\text{coll} \approx 10^{-9}\text{s} , \quad
T_\text{prec} \approx 190\text{d} , \quad
T \approx 1\text{d} .
\]
The precession is now entirely due to planet oblateness ($J_2=3\cdot 10^{-3}$). We now estimate the deflection angle due to scattering. When the scattering at impact parameter $b$ is integrated over all values,  for a path length of $2\pi a$ we find that
\[
\frac{\text{deflection}}{\text{orbit}} = \underbrace{\frac{N}{2\pi Ia^3}}_{\text{density}} \cdot \int\limits_0^s \!\! \underbrace{\frac{2am}{bM\epsilon^2}}_{\text{deflection}} \cdot \underbrace{2\pi a\ 2\pi b\mathrm db}_{\text{volume shell}} =
\frac{4\pi N}{I\epsilon^2} \bigg(\frac{m}{M}\bigg)^{7/5}
.
\]
The orbital eccentricity $\epsilon$ accounts from the fact that the relative velocity is roughly $\sqrt{GM/a}\epsilon$, which becomes small for orbits with the same sense of rotation. Although there are many close encounters where the mutual gravity takes over the central force, the (very crude) estimate of the deflection angle is about $10^{-5}$ and $10^{-7}$ per orbit for the inner Solar System and the Uranus ring system, respectively, mainly due to high relative velocities.

\begin{figure}
\centering
\includegraphics{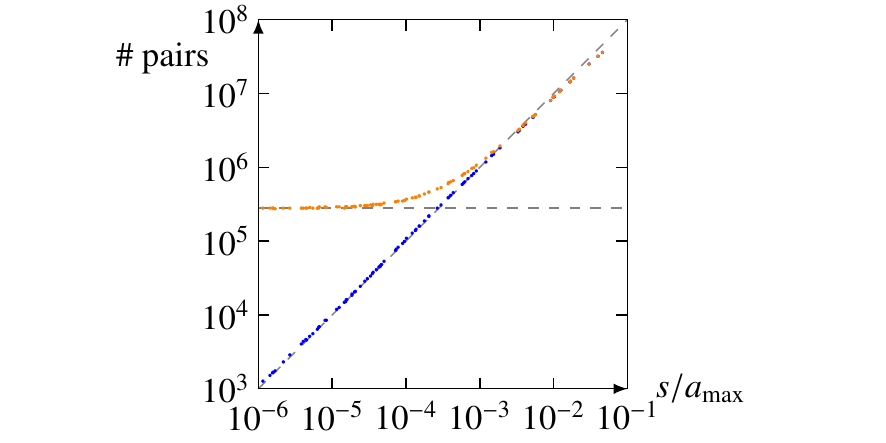}
\caption{Pairs vs.\ particle radius $s$ for a homogeneous disk with $a\leq a_\text{max}=2\text{au}$, $I\leq 10^{-3}$, $\epsilon\leq 10^{-3}$, and with $N=10^4$, from \citet{Aliberti2022github,Aliberti2022}.
\emph{Orange}: Twice the number of pairs
that needed to be checked in the apoapsis/periapsis filter. \emph{Blue}: Actual number of collision possibilities. The guideline with slope $1$ (\emph{dashed}) shows the approximate linear dependence of the collision pairs on $s$. The influence of the planet size can be seen in the apoapsis/periapsis filter for large $s$. Because there are two collision possibilities for each pair, the blue dots lie below or on the orange dots.}
\label{Figure3}
\end{figure}

\begin{figure}
\centering
\includegraphics{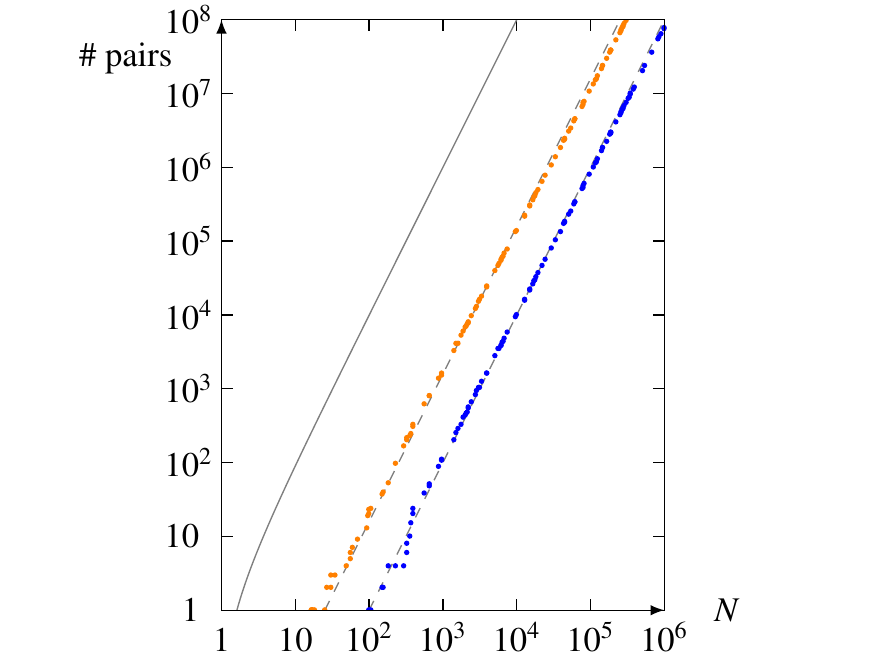}
\caption{Pairs vs.\ particle count $N$, as in Fig.~\ref{Figure3}, but with particle radius $s=2\cdot 10^{-3}\text{au}$. \emph{Orange}: Number of pairs that needed to be checked in the apoapsis/periapsis filter. \emph{Blue}: Actual number of collision pairs (that could ultimately collide). These determine the runtime and memory for the creation of the collision list; see Table~\ref{table2}. \emph{Solid} lines: $N(N-1)/2$. \emph{Dashed} line: guideline with a slope of $2$.
}
\label{Figure4}
\end{figure}

The algorithm maintains a list of all particle pairs that are on a collision course. As any pair can only collide near the line of intersection (in Fig.~\ref{Figure2}) of the two orbital planes, any random pair has the probability $\approx 2s/a$ of being on a collision trajectory. An estimate of the number of pairs on a collision course, or ``collision pairs'', is therefore $N^2s/a$. The results of simulations shown in Figs.~\ref{Figure3}-\ref{Figure4} validate these estimates.

As the algorithm steps from one collision to the next, the (average) time-step is equal to the collision time $T_\text{coll}$. Although exact precision is already lost in one orbit if corrections for secular motion are not included, we expect collision detection using the Kepler orbits to be able to give reliable statistical results for $T/N<T_\text{coll}<T_\text{prec}$.

\begin{table}[tb]
\caption{Algorithmic efficiency}
\label{table2}
\centering
\begin{tabular}{l|ll}
\hline
\hline
algorithmic step & runtime & memory \\
& $\propto O(\cdot)$ & $\propto O(\cdot)$ \\
\hline
create particle list & $N$ & $N$ \\
sort particle list & $N\log N$ & $1$ \\
create collision list &  $N^2\epsilon$ & $\dfrac{N^2 s}{a}$ \\
sort collision list &  $\dfrac{N^2s}{a}\log\dfrac{N^2s}{a}$ & $1$ \\
\hline
reduce particle list & $\log N$ & $0$ \\
create $\kappa$ fragments & $\kappa$ & $\kappa$ \\
sort $\kappa$ fragments & $\kappa\log\kappa$ & $1$ \\
merge particle lists & $N+\kappa$ & $1$ \\
reduce collision list & $\dfrac{Ns}{a}\log\dfrac{N^2s}{a}$ & $0$ \\
create new collision list & $\kappa N$ & $\dfrac{\kappa Ns}{a}$ \\
merge collision lists & $\dfrac{N^2s}{a}+\dfrac{\kappa Ns}{a}$ & $1$ \\
\hline
total simulation & $N^2\epsilon+\dfrac{N^4s^3}{Ia^3}$ & $\dfrac{N^2 s}{a}+\dfrac{N^4 s^3}{Ia^3}$ \\
\hline
\hline
\end{tabular}
\tablefoot{Order estimations, in big $O$, of the time and memory requirement in Keplerian collision detection. The top shows the initialisation, the middle rows show  the managing of one collision, and the bottom is for the full simulation. This total is found because there are an expected $O(N^2s^2/Ia^2)$ collisions in any fixed simulation time. For comparison, the time complexity for tree codes scales as $O(N^{4/3}I^{-1/3}\log N)$. The parameter $\kappa$ is the number of fragments produced in a collision.
}
\end{table}

\begin{figure}
\centering
\includegraphics{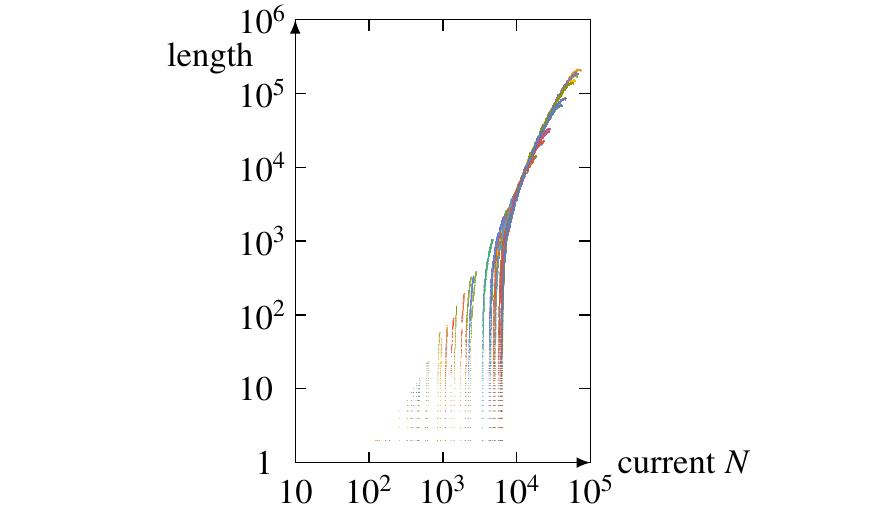}
\caption{Length of the list of collision pairs vs.\ particle count (same parameters as in Fig.~\ref{Figure4}). Only mergers are simulated, and consequently the $N$ value decreases by one at every step. The graph shows 64 runs.}
\label{Figure6}
\end{figure}

\subsection{Comparing time-steps}
We are interested in comparing this approach with numerical integration propagators with collision detection. In order to detect a collision in numerical integration, one must find nearest neighbors. In the tree code, one uses boxes of volume $\mathrm dx^3$ of a sufficiently small size to contain only one or a few particles. In one time-step, $\mathrm dt$, the change in position should not move the particle too many boxes away from its original position; otherwise, it becomes impossible to select neighbors. Alternatively, in an algorithm that uses a sorted list of the coordinates (so-called spatial hashing codes), a particle coordinate, say $x$, can overtake the values of other particles when its position changes by $\mathrm dx$. In this latter case, the number of particles that one particle overtakes in one step should also remain small in order to limit the number of neighboring particles that need to be inspected. As a result, numerical integration with collision detection requires not only several time-steps for one orbital period ($\mathrm dt \lessapprox T$) but also spatial steps of the order of the inter-particle distance ($\mathrm dx \lessapprox \overline{r_{ij}}$). We may estimate the average distance by assuming a homogeneous distribution of the particles. We then find for a disk with $a=4\text{au}$, $I=.1$:
\[
\overline{r_{ij}} = a\Gamma(\tfrac{4}{3}) \bigg(\frac{3I}{2N}\bigg)^{1/3} = \frac{2\text{au}}{N^{1/3}}
, \quad
\mathrm dt \lessapprox \frac{T\Gamma(\tfrac{4}{3})}{2\pi} \bigg(\frac{3I}{2N}\bigg)^{1/3}  = \frac{220\text{d}}{N^{1/3}}
.
\]
Clearly, the time-steps for numerical integration actually need to be quite small.

\begin{figure}
\centering
\includegraphics{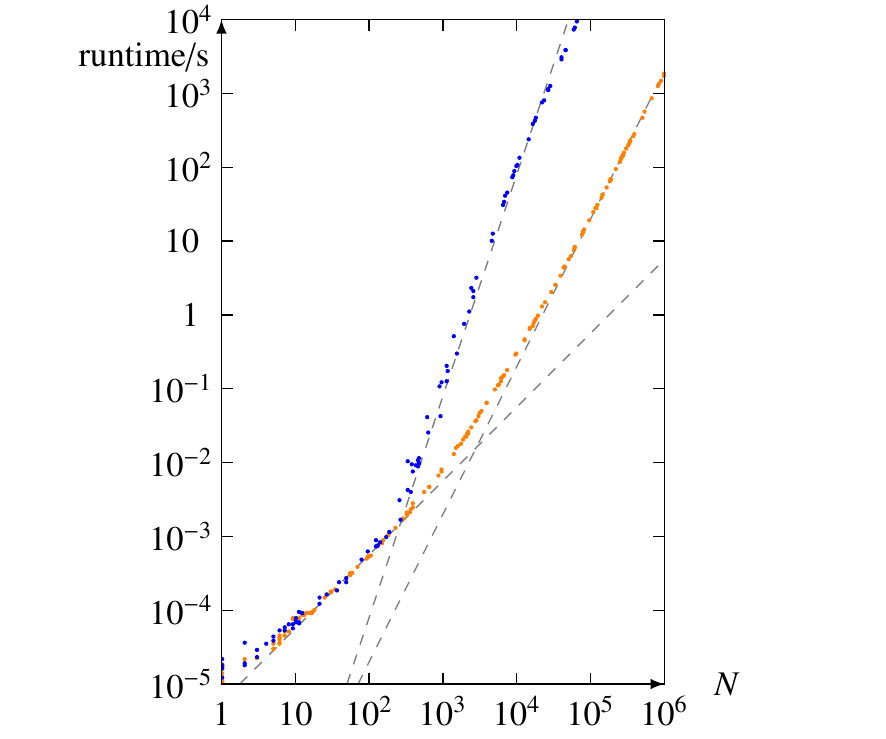}
\caption{
Initialisation- (\emph{orange}) and simulation runtime (\emph{blue}) vs.\ particle count $N$, as in Fig.~\ref{Figure4}. Dashed lines are  guidelines with slopes of 1, 2, and 3. The initialization time is respectively linear and quadratic in $N$ for $N<1/\epsilon$ and $N>1/\epsilon$, in accordance with the estimates in Table~\ref{table2}. However, because only mergers were simulated, there were at most $N$ collisions, which results in a runtime of $O(N^3)$ instead of $O(N^4)$.
}
\label{Figure5}
\end{figure}

\subsection{Efficiency of the algorithm}
Apart from the advantage of having large time-steps, another benefit of our method is that only the collision products need to be tested for possible future collisions with the set of existing particles. This involves $\propto N$ computations per collision. However, a list of collision possibilities needs to be maintained. The length of this list is expected to be of the order $N^2s/a$. This list takes up memory and therefore requires careful manipulation. At the creation or the removal of a particle, an average number of $2Ns/a$ new collision possibilities is added or removed, respectively. Hence, after $N/2$ collisions the list has mostly changed. This is understandable, because then most particles are replaced by new particles. It also means that for $Ns/a\gg 1$, most possibilities do not actually happen. Figure~\ref{Figure4} shows the initial list length and Fig.~\ref{Figure6} shows how this length changes during the simulation.

\begin{figure}[t]
\centering
\includegraphics{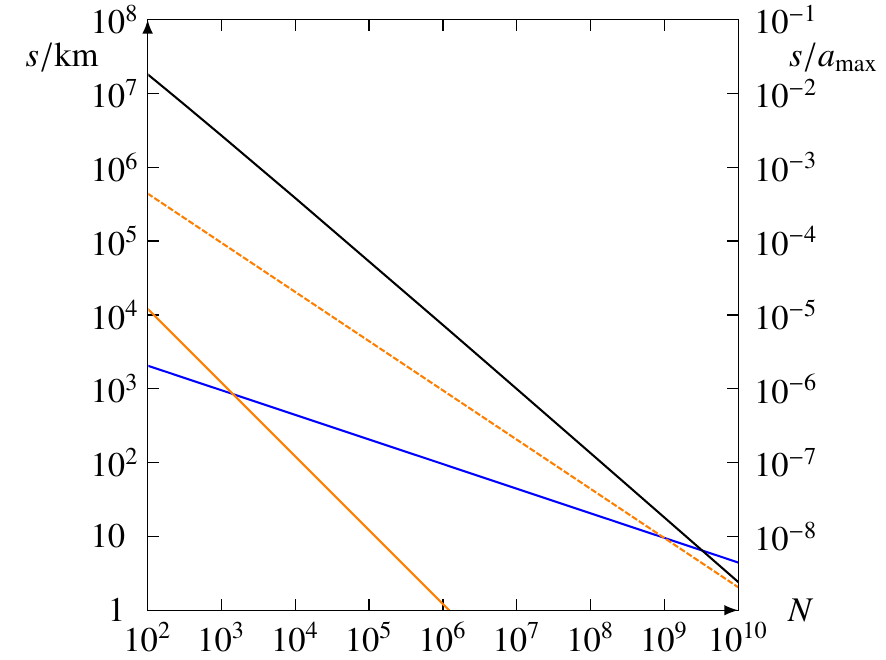}
\caption{Particle radius $s$ vs.\ particle count $N$ for a disk system with $I\leq .1$ and $a\leq 4\text{au}$. Above the \emph{black} line, the numerical integration propagator with tree code for collision detection beats the analytic propagator using Kepler orbits (in terms of time efficiency). Precession due to a Jupiter has a small effect above the \emph{orange} line, where $T_\text{coll}=T_\text{prec}$, and has a negligible effect above the \emph{orange dashed} line, $aT_\text{coll}=sT_\text{prec}$.
If the mass of three Earths is distributed over equal-sized planetoids (of Earth density), one is constrained
to the \emph{blue} line. On this line, precession is small for $N>10^3$ but only attains the much smaller error of $s$ per orbit for $N>10^9$. Clearly, tree-code is better for these high number densities.}
\label{Figure7}
\end{figure}

Table~\ref{table2} sums up the efficiencies in the various steps of the algorithm. 
Figure~\ref{Figure5} shows the measured runtime for a large set of simulations with different particle numbers. As we did not include defragmentation but only mergers,  Figure~\ref{Figure7} shows the values of particle radius $s$ and particle count $N$ for which the Kepler collision detection is more efficient than the algorithmic efficiency $(T_\text{sim}/\mathrm dt)N\log N\propto N^{4/3}\log N$ of numerical integration with nearest neighbor search using a tree code.

\begin{figure}[t]
\centering
\includegraphics{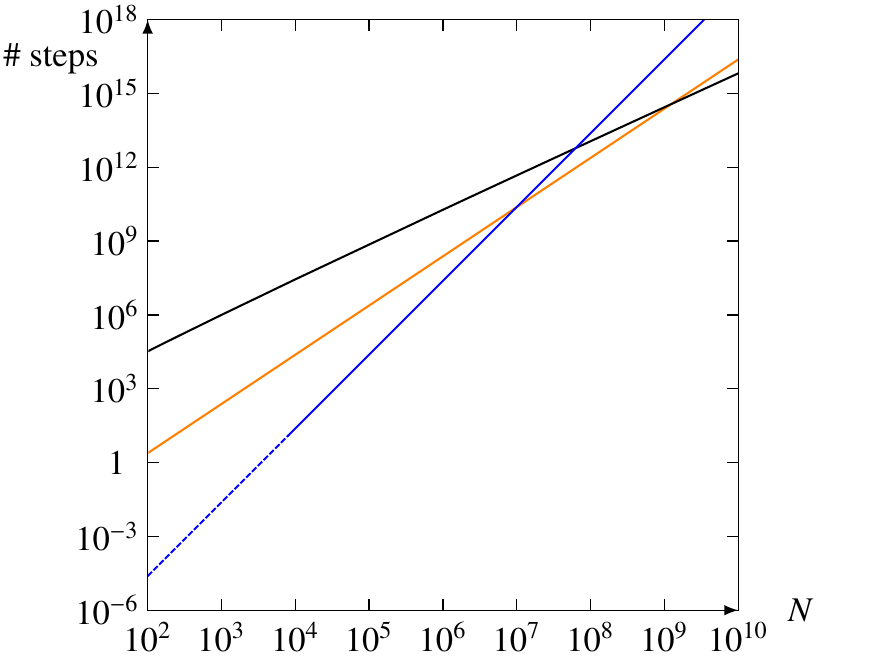}
\caption{Theoretical algorithmic efficiency: number of computational steps (big $O$) for one orbital period vs.\ particle count $N$. \emph{Blue}: Intersecting Kepler orbits without secular dynamics. \emph{Orange}: with secular dynamics. The \emph{blue dotted} line indicates where precession due to secular dynamics cannot be neglected. The \emph{black} line shows the estimated number of steps for a tree code and/or spatial hashing.
The slopes of the lines are $3$, 
$2$, and $4/3$, respectively.}
\label{Figure8}
\end{figure}

In order to make another comparison between the methods, consider a Solar System with a fixed amount of material volume. We take a disk with $a=4\text{au}$, $I=.1$, and a mass  $Nm=
3M_\text{Earth}$ and with particles of the density of Earth. We thus have $Ns^3/a^3=10^{-15}$. The resulting average $s$ is also plotted in Fig.~\ref{Figure7}. The efficiencies for collision detection for the cases without and with precession are then $(N^2s/a)(T/T_\text{coll})\propto N^3$ and $N^2T/T_\text{prec}\propto N^2$, respectively. These are shown in Fig.~\ref{Figure8}, and are also compared with the algorithmic efficiency numerical integration with a tree code. For a Solar System model with these parameter values, the octree code is faster for $N>10^8$.

\subsection{Applications and  neglected effects}
\label{Sec2.3}
We can think of the following applications:
\begin{inparaenum}[(i)]
\item Gravity assists at planetary flybys of a probe traversing the Solar System: the collision detection scheme calculates the times of passage using the sphere of influence.
\item The tracking of a comet as its orbit is perturbed at close encounters by the planets.
\item The rings of Saturn, with contact collisions and/or the gravitational influence of shepherd moons.
\item Growth by merging planetesimals into protoplanets, in young planetary systems.
\item A simple model to study the Kessler syndrome, where artificial satellites break apart due to impacting space debris.
\end{inparaenum}

Because of the approximations, collision detection with Kepler orbits is often inaccurate. However, sometimes accuracy is not the aim, or not even possible due to the chaotic nature of the problem. Instead, we may simply want to find out what could happen, and calculate the probabilities of the various outcomes.
The speedup allows sampling of initial states, either by adding many small perturbations to one initial state or by adding one perturbing body with many initial states from a large phase-space volume. Analytic propagation with collision detection based on the Kepler orbits neglects the following effects:
\begin{inparaenum}[(i)]
\item orbital precession due to mutual gravity or oblateness of the central body (as discussed in Sect.~\ref{Sec6.3}),
\item three-body gravitational scattering,
\item planetary migration due to interaction with the gas in the protoplanetary disk,
\item capture of planets in mean-motion resonances, 
\item the Kozai mechanism,
\item moons and binaries,
\item atmospheric and
\item tidal drag, 
\item solar wind, and
\item  the Poynting-Robertson effect.
\end{inparaenum}

\section{The algorithm}
\label{Sec3}
\subsection{Initialization}
We have a system of planets, or particles orbiting a central mass. The particles are numbered $j=1,\ldots,N$. For each particle, we store the following set of variables:
\begin{equation}
\big\{ t^0_j, \quad a_j, \quad c_j, \quad s_j, \quad m_j, \quad \boldsymbol r^0_j, \quad \boldsymbol L_j, \quad \boldsymbol\epsilon_j, \quad \omega_j \big\}
.
\label{parameters}
\end{equation}
Here, $t^0$ is the time of creation, $a$, $c$ are the orbital radius and focal distance, $s$ is the particle radius, $m$ is the particle mass, $\boldsymbol r^0$ is its position, $\boldsymbol L$ its angular momentum vector, $\boldsymbol\epsilon$ is the eccentricity vector, and $\omega$ is the frequency.

An initial state would consist of many particles in nearly circular orbits and therefore with small $\epsilon$. If one draws random numbers for the mean anomaly from the interval $[0,2\pi]$, the resulting (smoothed) phase-space distribution will become stationary; if the values of $\varpi$, $\ascnode$ are sampled from $[0,2\pi]$, the distribution will become axisymmetric; if also $\cos I$ is drawn from $[0,1]$, it will become spherically symmetric \citep[see][]{Savransky2011}. To simulate a thin disk, one takes small values for $I$. Equations (\ref{Lvec}) and (\ref{epsvec}) give the vectors $\boldsymbol L$ and $\boldsymbol\epsilon$ in terms of $a$ and $\epsilon$ and the angles $I$, $\varpi$, and $\ascnode$.

We then sort the particle list by increasing value of periapsis. This will allow the implementation of the apoapsis/periapsis filter \citep{Baraff1992}. The next step is to consider all particle pairs, and list the pairs that can collide. For these, we also store the calculated collision time ${t^1}_{(i,j)}$. If sufficient memory is available, it is possible to store the parameters in Eq.~(\ref{parameters}) for the new particle that would be formed after the collision. The soonest collision is at the top of the list.

\subsection{Main loop}
\begin{enumerate}
\item
If the collision list is empty, end the simulation.
\item
Take the pair $(i,j)$ with the soonest collision: the first in the list. 
\item
Update the time $t$ to the time $t^1_{(i,j)}$ of the collision.
\item
Remove any pair containing $i$ and any pair containing $j$ from the pairs list.
\item
Remove the particles $i$ and $j$   from the particle list.
\item
If the orbit of the new particle intersects the central mass or is unbound, go to the next collision on the list.
\item
Create new particle(s) defined by 
$\big\{t^1_{(i,j)}, a, c, s, m, \boldsymbol r^0, \boldsymbol L, \boldsymbol\epsilon, \omega\big\}$.
\item
For any new particle, consider the other particles and decide if the pair is on a collision course. If this is the case, calculate the time of the earliest collision.
\item
Make a sorted list of the new collision possibilities with a record of the collision time and the pair, soonest collision first.
\item 
Merge this sorted list with the existing sorted list of collision possibilities into a full list of pairs, sorted by time of the collision event, soonest collision first.
\end{enumerate}

\subsection{Determining if a pair is on a collision course}
At the initialization,
pairs of particles need to be considered for a possible future collision. Also, during the simulation, each time a new particle is created, all existing particles need to be paired with the new particle and considered for a possible future collision. However, as we implement the sweep and prune method, only pairs need to be considered with an overlap in the range of radial motion. The radial coordinate for each particle $i$ ranges over the interval from the periapsis to the apoapsis:
\[
\big[a_i-c_i-s_i,a_i+c_i+s_i\big] ,
\]
including an extension $s_i$ of the size of the particle, or with the substitution of its gravitational reach $s_i=(m_i/M)^{2/5}a_i$. It is sufficient to compare each particle $i$ with the particles $j=i+1, j=i+2, \ldots$. Because the list is sorted, we have $a_i-c_i-s_i\leq a_j-c_j-s_j$. As long as $a_i+c_i+s_i\geq a_j-c_j-s_j$, the intervals for $i$ and $j$ overlap and the pair is a candidate for collision. Once we encounter the first $j$ where $a_i+c_i+s_i<a_j-c_j-s_j$, there are no more particles that can interact with $i$ and we can go to particle $i+1$. If $\epsilon$ is the average eccentricity, only a fraction of $2\epsilon$ of the total number of pairs $N^2/2$ need to be checked. The resulting reduced number of checks in our numerical simulations is shown in Figs.~\ref{Figure4} and \ref{Figure5}.

For a particle $i$ created during the simulation, the selection of pairs is slightly different. Again, it is sufficient to consider only particles $j$ with periapsis smaller than the apoapsis of $i$.
However, this time we have to start at $j=1$.
\begin{enumerate}
\item
Consider a pair, say $(i,j)=(1,2)$; we assume that $a_1>a_2$. First we need to
find the minimal orbit intersection distance to decide whether or not there can be a collision. For this, retrieve $m_1, \boldsymbol L_1, \boldsymbol\epsilon_1$, and $m_2, \boldsymbol L_2, \boldsymbol\epsilon_2$. Then calculate the direction of the nodal line (see Fig.~\ref{Figure2}), the semi-latus recti, the intersection points, and the velocities at these points:
\begin{align}
\boldsymbol K & = \boldsymbol L_1 \times \boldsymbol L_2 , && \nonumber \\
K &= \sqrt{\boldsymbol K\bullet \boldsymbol K} , && \nonumber\\
\ell_1 &= \frac{\boldsymbol L_1\bullet\boldsymbol L_1}{GMm_1^2} , &
\ell_2 &= \frac{\boldsymbol L_2\bullet\boldsymbol L_2}{GMm_2^2} , \nonumber\\
\boldsymbol r_1 &= \frac{\boldsymbol K\ell_1}{\pm K + \boldsymbol\epsilon_1\bullet\boldsymbol K} , &
\boldsymbol r_2 &= \frac{\boldsymbol K\ell_2}{\pm K + \boldsymbol\epsilon_2\bullet\boldsymbol K} ,
\label{lineintersect} \\
r_1 &= \sqrt{\boldsymbol r_1\bullet\boldsymbol r_1} , &
r_2 &= \sqrt{\boldsymbol r_2\bullet\boldsymbol r_2} , \nonumber \\
\boldsymbol v_1 &= \frac{\boldsymbol L_1}{m_1\ell_1} \times \bigg( \boldsymbol\epsilon_1  + \frac{\boldsymbol r_1}{r_1} \bigg) , &
\boldsymbol v_2 &= \frac{\boldsymbol L_2}{m_2\ell_2} \times \bigg( \boldsymbol\epsilon_2  + \frac{\boldsymbol r_2}{r_2} \bigg)
.
\label{velocities}
\end{align}
We write $G$ for Newton's constant. Equation~(\ref{lineintersect}) for the intersection points is derived in Sect.~\ref{Sec4.1}. Equation~(\ref{velocities}) is Eq.~(\ref{decomp}) from Appendix~\ref{AppB}. The two pairs $(\boldsymbol r_1,\boldsymbol r_2)$ at the opposite sides are indicated with the plus/minus symbol. The steps that now follow must be performed on both of the pairs.

\item
Next, calculate for both particles the (approximate) points on the orbits where the distance is minimal:
\begin{align}
\boldsymbol d &= \boldsymbol r_2 - \boldsymbol r_1 , &&
\nonumber \\
\boldsymbol w & = \boldsymbol v_1 \times \boldsymbol v_2 , &&\nonumber \\
(w^2) &= \boldsymbol w\bullet \boldsymbol w , &&
\nonumber \\
\boldsymbol r'_1 &= \boldsymbol r_1 + \bigg( \boldsymbol d \bullet \frac{\boldsymbol v_2 \times \boldsymbol w}{(w^2)} \bigg) \boldsymbol v_1 , &
\boldsymbol r'_2 &= \boldsymbol r_2 + \bigg( \boldsymbol d \bullet \frac{\boldsymbol v_1 \times \boldsymbol w}{(w^2)} \bigg) \boldsymbol v_2 ,
\label{minimum} \\
\boldsymbol r_1 &= \boldsymbol r'_1 , &
\boldsymbol r_2 &= \boldsymbol r'_2 ,
\nonumber \\
\boldsymbol d &= \boldsymbol r_2 - \boldsymbol r_1 , &&
\nonumber \\
d &= \sqrt{\boldsymbol d\bullet\boldsymbol d} . &&
\nonumber
\end{align}
Equation~(\ref{minimum}) for the collision points $\boldsymbol r'_1$ and $\boldsymbol r'_2$ is derived in Sect.~\ref{Sec4.2}.

\item
Retrieve $a_1$, $s_1$ and $a_2$, $s_2$. Decide whether or not an interaction can take place.
\begin{align*}
d &< s_1+s_2  & \Longrightarrow & \quad
\text{1 and 2 collide.} \\
d &< (m_1/M)^{2/5}a_1 & \Longrightarrow & \quad
\text{1 can perturb orbit 2.} \\
d &< (m_2/M)^{2/5}a_2 & \Longrightarrow & \quad
\text{2 can perturb orbit 1.}
\end{align*}
If there is no contact or perturbation:
go to the next pair.
\end{enumerate}

\subsection{Deterministic collision time}
We now give the steps in the calculation of the exact collision moment.
\begin{enumerate}
\item
Retrieve $t^0_1$, $\omega_1$, $\vec r^0_1$ and $t^0_2$, $\omega_2$, $\vec r^0_2$ of the particles involved (with $\omega_1<\omega_2$). Calculate the first time a particle passes the crossing point in Fig.~\ref{Figure2}. These times are denoted by $t^1_1$, $t^1_2$.
\begin{align}
(\epsilon_1^2) &= \boldsymbol \epsilon_1\bullet\boldsymbol\epsilon_1 , \nonumber \\
(\epsilon_2^2) &= \boldsymbol \epsilon_2\bullet\boldsymbol\epsilon_2 , \nonumber \\
r_1 &= \sqrt{\boldsymbol r_1\bullet\boldsymbol r_1} , \nonumber \\
r_2 &= \sqrt{\boldsymbol r_2\bullet\boldsymbol r_2} , \nonumber \\
z &=
\bigg(\frac{\vec r_1}{a_1} - \frac{\mathrm ir_1\vec v_1}{a_1^2\omega_1}\bigg) \bullet \bigg( \frac{\vec r^0_1-\vec\epsilon_1\ \vec\epsilon_1\bullet \vec r^0_1}{a_1-(\epsilon_1^2)a_1} + \vec\epsilon_1 \bigg) + \frac{\vec r^0_1\bullet\vec\epsilon_1}{a_1} + (\epsilon_1^2)
\label{E1E2} \\
\Delta E_1 &= \arg z , \quad 0 \leq \Delta E_1 < 2\pi ,
\nonumber \\
z &=
\bigg(\frac{\vec r_2}{a_2} - \frac{\mathrm ir_2\vec v_2}{a_2^2\omega_2}\bigg) \bullet \bigg( \frac{\vec r^0_2-\vec\epsilon_2\ \vec\epsilon_2\bullet \vec r^0_2}{a_2-(\epsilon_2^2)a_2} + \vec\epsilon_2 \bigg) + \frac{\vec r^0_2\bullet\vec\epsilon_2}{a_2} + (\epsilon_2^2)
\nonumber \\
\Delta E_2 &= \arg z , \quad 0 \leq \Delta E_2 < 2\pi ,
\nonumber \\
t^1_1 &= t^0_1 + \frac{\Delta E_1}{\omega_1} - \frac{\vec\epsilon_1 \times (\vec r_1-\vec r^0_1)}{1-(\epsilon_1^2)} \bullet \frac{\vec L_1}{GMm_1} ,
\label{passage} \\
t^1_2 &= t^0_2 + \frac{\Delta E_2}{\omega_2} - \frac{\vec\epsilon_2 \times (\vec r_2-\vec r^0_2)}{1-(\epsilon_2^2)} \bullet \frac{\vec L_2}{GMm_2}
.
\nonumber
\end{align}
These equations are derived in Sect.~\ref{Sec5}. Equation~(\ref{E1E2}) is obtained by combining Eqs.~(\ref{cos}-\ref{sin}). The complex number has unit modulus. Equation~(\ref{passage}) for the passage times follows from Eq.~(\ref{deltat}).

\item
Evaluate the small dimensionless parameter.
\begin{align}
\boldsymbol u &= \boldsymbol v_2 - \boldsymbol v_1 , \nonumber \\
\delta &= \frac{1}{|t^1_1-t^1_2|} \textstyle\sqrt{\displaystyle \boldsymbol u\bullet\boldsymbol u \dfrac{(s_1+s_2)^2 - d^2}{(w^2)} }
.
\label{delta}
\end{align}
\item
Next calculate the exact number of periods $k$ 
that planet~1 
makes before colliding with planet~2. The method uses the continued fraction of the $\omega_2/\omega_1$. Initialize the loop:
\begin{align*}
q_0 &= \frac{2\pi}{\omega_1|t^1_1-t^1_2|} , & k_0 &= 1 , \\ 
q_1 &= \frac{2\pi}{\omega_2|t^1_1-t^1_2|} , & k_1 &= 0 
.
\end{align*}

\item
Start loop counter at $n=0$. The loop creates integer sequences $\alpha_n, k_n$ 
and the positive sequence $q_n$. These are the \emph{digits}, the denominators of the \emph{convergents}, and the \emph{remainders} of the continued fraction (we do not need the numerators, denoted $l_n$). The loop performs the iterations
\begin{align*}
\alpha_{2n} = \biggl\lfloor\frac{q_{2n}}{q_{2n+1}}\biggr\rfloor , \quad
q_{2n+2} &= q_{2n} - \alpha_{2n}q_{2n+1} , \\
\text{if}\quad q_{2n+2} &= 0 \quad\Longrightarrow\quad \text{next pair} \\
k_{2n+2} &= k_{2n} - \alpha_{2n} k_{2n+1} , \\
\alpha_{2n+1} = \biggl\lfloor\frac{q_{2n+1}}{q_{2n+2}}\biggr\rfloor , \quad
q_{2n+3} &= q_{2n+1} - \alpha_{2n+1}q_{2n+2} , \\
\text{if}\quad q_{2n+3} &= 0 \quad\Longrightarrow\quad \text{next pair} \\
k_{2n+3} &= k_{2n+1} - \alpha_{2n+1} k_{2n+2} . \\
\end{align*}
We use the notation $\lfloor\cdot\rfloor$ and $\lceil\cdot\rceil$ for the floor and the ceiling function. The time $T_\text{sim}$ to be simulated sets an upper bound for the solution:
\[
x_\text{max} = \frac{(T_\text{sim}-t^1_1)\frac{\omega_1}{2\pi}q_{2n+1}-(1+\delta)k_{2n+1}}{k_{2n}q_{2n+1}-q_{2n}k_{2n+1}}
.
\]
We then test the points with coordinates
\begin{align*}
x = \biggl\lceil \frac{1-\delta}{q_{2n}} \biggr\rceil, \ldots ,
\biggl\lfloor\frac{1+\delta}{q_{2n+2}}\biggr\rfloor
, & \\
\text{if}\quad x &> x_\text{max} \quad\Longrightarrow\quad \text{next pair} \\
y &= \text{max}\bigg( 0, \biggl\lceil \frac{q_{2n}x-1-\delta}{q_{2n+1}} \biggr\rceil \bigg)
.
\end{align*}
If for any of these points $1-\delta < xq_{2n}-yq_{2n+1}$, 
then we have found the solution. If not, we increase $n$ by $1$ and check this next.

\item
The solution is
\begin{equation}
k = xk_{2n} - yk_{2n+1} , \quad
t^0 = 
t^1_1 + \frac{2\pi k}{\omega_1}
.
\label{deterministictime}
\end{equation}
Equation~(\ref{deterministictime}) for the deterministic collision time is derived in Sect.~\ref{Sec5.1}.
\end{enumerate}
The solution $k=0$ means that the pair $(1,2)$ is about to collide within the present period. If a gravitational scattering between the same pair $(1,2)$ has happened at the previous time-step, 
the solution $k=0$ corresponds to the scattering that was just simulated, and therefore is invalid. However, for a different pair where one particle participated in this last scattering, the solution $k=0$ is valid, as it implies an immediate collision with a third particle. For three (or more) bodies inside each others sphere of influence, the multi-body gravitational scattering will therefore be treated as three (or more) successive two-body interactions.

\subsection{Stochastic collision time}
\label{Sec3.5}
Although the time of collision between two bodies is uniquely determined by the initial conditions, it is highly sensitive to the precise values of the creation times and the orbital periods.
Therefore, if the physical or numerical error in the (initial) values is bigger than $s/a$, the collision time becomes unpredictable.

If the collision process is assumed to be stochastic, one may adopt the following Monte Carlo method. First, a random number $\xi$ is drawn from the interval $[0,1]$. The moment of collision is calculated with
\begin{align}
\boldsymbol u &= \boldsymbol v_2 - \boldsymbol v_1 , \nonumber \\
t^0 &= t_1^1 + \frac{(-\log\xi)(2\pi)^2}{2GM} \textstyle\sqrt{\dfrac{(w^2)}{\boldsymbol u\bullet\boldsymbol u} \dfrac{a_1^3a_2^3}{(s_1+s_2)^2 - d^2} }
.
\label{stochastictime}
\end{align}
Equation~(\ref{stochastictime}) is derived in Sect.~\ref{Sec5.2}.

\begin{figure}
\centering
\includegraphics{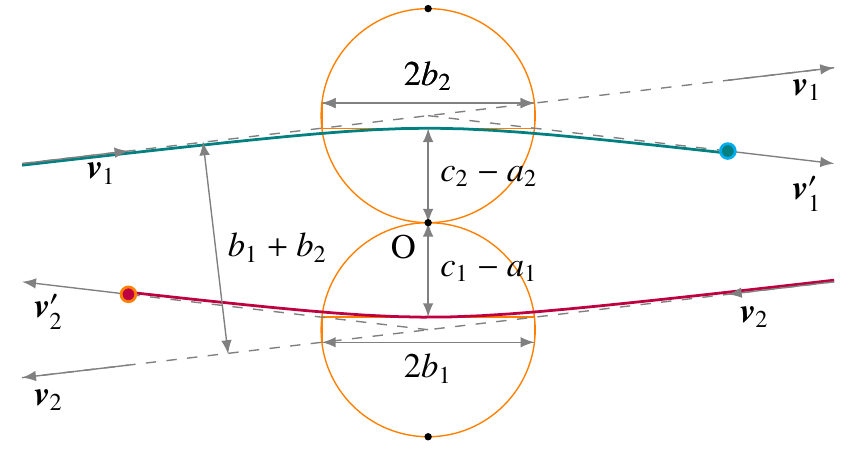}
\caption{Gravitational scattering between particles 1 and 2 in the center-of-mass frame. The initial velocities $\boldsymbol v_1$, $\boldsymbol v_2$ are scattered in the directions $\boldsymbol v'_1$, $\boldsymbol v'_2$ along the asymptotic lines (\emph{dotted}). The actual orbits are hyperbolas (\emph{blue}, \emph{purple}). The focal points (\emph{black} dots), with O being the common focal point of the orbits, lie on the \emph{orange} circles. The four points of intersection of a circle with the asymptotes form rectangles with dimensions of the transverse axis $2a_j$ by the conjugate axis $2b_j$ \citep{Adams2021}. The impact parameter is the sum $b=b_1+b_2$.}
\label{Figure9}
\end{figure}

\subsection{The collision}

\begin{figure}
\centering
\includegraphics{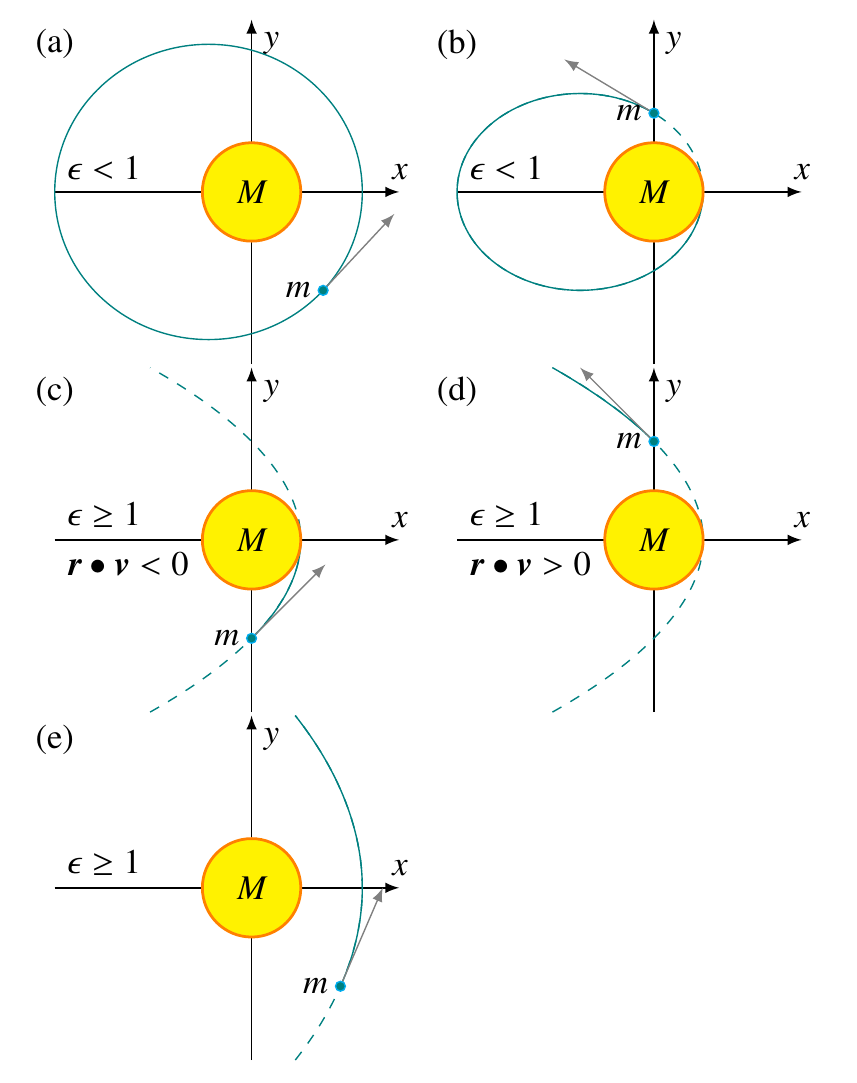}
\caption{Five cases for the obit of a collision fragment. 
(a) The particle is bound. (b), (c) The particle collides with the central mass. (d), (e) The particle leaves the system. For long intervals between collisions ($T_\text{coll}\gg T$), cases (b-e) can be dealt with by removing the particle. 
Otherwise, the fragments need to be stored and paired in a similar manner to the bound particles.}
\label{Figure10}
\end{figure}

In order to simulate the collision or scattering event, a physical model of the merger or break-up of the particles needs to be implemented. In the case of pure gravitational scattering (close passage), one can use the formulas from Appendix~\ref{AppC} for the momentum exchange. This elastic collision is depicted in Fig.~\ref{Figure9}. We now outline how to find the orbit of the new particle in case of a merger between particles 1 and 2.
\begin{enumerate}
\item
For a simple merger, the new particle has radius, mass, position, velocity, and angular momentum\footnote{
As pointed out by the referee, orbital angular momentum is not strictly conserved, because it is transferred into spin for oblique collisions:
this spin becomes
$\boldsymbol S=\boldsymbol S_1+\boldsymbol S_2 + (m_1m_2/m) \boldsymbol d\times \boldsymbol u$,
provided $|\boldsymbol S|<(2/5)\sqrt{Gm^3s}\!$.
} calculated from the basic conservation laws:
\begin{align*}
s &= (s_1^3+s_2^3)^{1/3} && \text{(material volume)}, \\
m &= m_1 + m_2 && \text{(mass)}, \\
\boldsymbol r &= \frac{m_1\boldsymbol r_1+m_2\boldsymbol r_2}{m} && \text{(center-of-mass motion)}, \\
\boldsymbol v &= \frac{m_1\boldsymbol v_1+m_2\boldsymbol v_2}{m} && \text{(momentum)}, \\
\boldsymbol L &= m\boldsymbol r \times \boldsymbol v 
&& \text{(angular momentum)}
.
\end{align*}
\item
Decide whether or not the new particle collides with the central body. We suppose that it is a sphere of radius $S$.
\begin{align*}
\ell &=  \frac{\boldsymbol L\bullet\boldsymbol L}{GMm^2} , \\
r & = \sqrt{\boldsymbol r\bullet\boldsymbol r} , \\
\boldsymbol\epsilon &= \frac{\boldsymbol v\times \boldsymbol L}{GMm} - \frac{\boldsymbol r}{r} , \\
\epsilon &= \sqrt{\boldsymbol\epsilon\bullet\boldsymbol\epsilon}
.
\end{align*}
In the general case that several collision fragments are created in the collision, there are five cases, with three outcomes (see Fig.~\ref{Figure10}) for a fragment. First, consider the cases where there is no crossing:
\begin{align*}
\ell > (1+\epsilon)S  \quad\text{and}\quad \epsilon &< 1 & \Longrightarrow & \quad \text{new particle stays.} \\
\ell > (1+\epsilon)S  \quad\text{and}\quad \epsilon &\geq 1 & \Longrightarrow & \quad \text{particle escapes.}
\end{align*}
In the remaining cases, $\ell\leq (1+\epsilon)S$ and the orbit crosses the central body.
\begin{align*}
\epsilon &\geq 1 \quad \text{and}\quad \boldsymbol r\bullet\boldsymbol v>0 & \Longrightarrow & \quad \text{particle escapes.} \\
\epsilon &\geq 1 \quad \text{and}\quad \boldsymbol r\bullet\boldsymbol v\leq 0 & \Longrightarrow & \quad \text{collides with $M$.} \\
\epsilon &< 1 & \Longrightarrow & \quad \text{collides with $M$.}
\end{align*}
Only in the first case does the particle stay in a bound orbit, and we continue; otherwise, the particle is removed. 

For a merger, there is only one collision product, and the newly formed particle will always have $\epsilon<1$ because (total) energy can only decrease. 

\item
Calculate the required orbital parameters of the new particle:
\[
a =  \frac{\ell}{1-\epsilon^2} , \quad
c = a\epsilon ,  \quad
\omega = \sqrt{\displaystyle\frac{GM}{a^3}}
.
\]
\item
Register $\big\{ t^0, a, c, s, m, \boldsymbol r, \boldsymbol L, \boldsymbol\epsilon, \omega \big\}$ of the new particle.
\item
Sort the list of new particles by increasing $a-c-s$.
\item
Merge these new-particle lists with the existing-particle list.
\end{enumerate}
{Go to the next time-step.}

\section{Calculating points of closest approach}
\label{sec4}
In this section, we derive an approximation for the points on two Kepler orbits with minimal separation. This distance is called the minimum orbit intersection distance (MOID). As we are considering possible collisions between planets, we are interested in the case where the MOID for the orbits of planet~1 and planet~2 is less than $s_1+s_2$. We assume that the planet radii $s_j$ are many orders of magnitude smaller than the orbital radii $a_j$. Then, for an angle $I$ between the orbital planes of larger than $(s_1+s_2)/(a_1+a_2)$, the MOID will be close to the line of intersection
of the orbital planes \citep{Hoots1984,Manley1998}. This principle is illustrated in Fig.~\ref{Figure2}: Outside a cylinder with radius $(s_1+s_2)/(2\sin\tfrac{I}{2})$ about the ``mutual nodal line'', all points in orbit~1 are separated by more than $s_1+s_2$ from points in orbit~2. Any collision must therefore happen inside the cylinder. The cylinder is only large for very small inclinations. If the system is a disk with an average inclination angle $\bar I$, these small inclinations are rare for $N<\bar Ia^2/s^2$. 

Because the range for gravitational scattering can be larger, our approach will only work for low-mass planets. The iterative scheme converging to the MOID that projects the points onto the orbit followed by linearization is described in \citet{Hoots1984}. 
{Various other methods to obtain the MOID have been found}
\citep[{see e.g.}][]{Gronchi2005,Milisavljevic2010,Segan2011,Wizniowski2013,Hedo2018}.

\subsection{Intersecting orbit~1 with orbital plane 2}
\label{Sec4.1}
\begin{figure}
\centering
\includegraphics{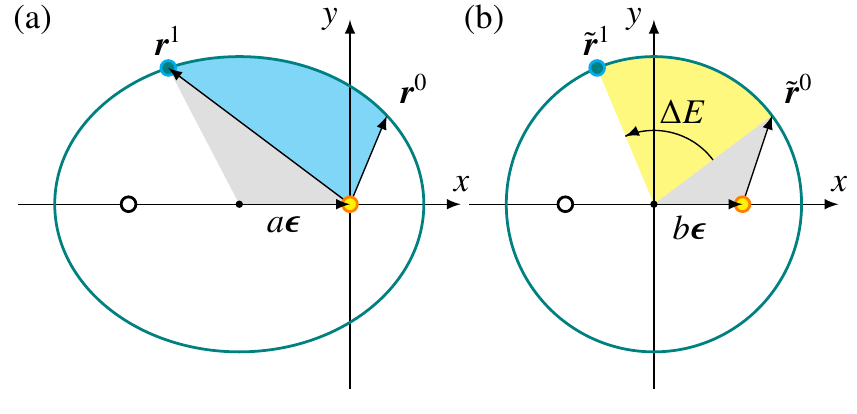}
\caption{Elliptical Kepler orbit (a) and orbit squeezed into a circle (b) from scaling the $x$-axis by $b/a$. The travel time from $\boldsymbol r^0$ to $\boldsymbol r^1$ is equal to the ratio of the swept (\emph{cyan}) area in (a) to $\pi ab$ times the period. In (b), this ratio is the (\emph{yellow}) circle segment plus the (\emph{gray}) triangle minus the image of the (\emph{gray}) triangle in (a) over $\pi b^2$. The circle segment has area $\Delta E b^2/2$, with $\Delta E=E^1-E^0$ being the difference in eccentric anomaly.}
\label{Figure11}
\end{figure}

The Kepler orbit of a planet is entirely determined by its angular momentum $\boldsymbol L$ and eccentricity vector $\boldsymbol\epsilon$ \citep{Goldstein1964}. The angular momentum is normal to the place of the orbit and the vector $a\boldsymbol\epsilon$ points from the center of the ellipse to the focal point where the central mass is located (see Fig.~\ref{Figure11}). Now let us consider two orbits, for planet~1 and planet~2, specified by $\boldsymbol L_1, \boldsymbol\epsilon_1$ and $\boldsymbol L_2, \boldsymbol\epsilon_2$, respectively. The line of intersection of the two orbital planes, or the nodal line, can be found as follows. Because the angular momenta $\boldsymbol L_1$ and $\boldsymbol L_2$ are both perpendicular to the nodal line, a direction vector 
of the nodal line is 
\[
\boldsymbol K = K\hat{\boldsymbol K} = \pm \boldsymbol L_1 \times \boldsymbol L_2
.
\]
The plus/minus symbol
indicates the two opposite directions in which the intersection points with an orbit are found. Because the eccentricity vector $\boldsymbol\epsilon$ of an orbit points from the central mass towards the periapsis, the true anomaly $\nu$ of the intersection point in the direction $\boldsymbol K$ is given by
\[
\cos \nu = \frac{\boldsymbol\epsilon\bullet\hat{\boldsymbol K}}{\epsilon}
.
\]
The point of intersection can now be found from the formula for the orbit Eq.~(\ref{orbit}). We find
\[
\boldsymbol r = r\hat{\boldsymbol K} =
\frac{(1-\epsilon^2)a}{1+\epsilon\cos\nu} \hat{\boldsymbol K} =
\frac{(1-\epsilon^2)a}{1+\boldsymbol\epsilon\bullet \hat{\boldsymbol K}} \hat{\boldsymbol K} =
\frac{(1-\epsilon^2)a}{K+\boldsymbol\epsilon\bullet\boldsymbol K} \boldsymbol K
.
\]
Therefore,  for the two pairs of intersection points, we obtain
\begin{align*}
\boldsymbol r_{1,\pm} &= \frac{(1-\epsilon_1^2)a_1 \boldsymbol L_1 \times \boldsymbol L_2}{\pm|\boldsymbol L_1 \times \boldsymbol L_2| + \boldsymbol\epsilon_1\bullet(\boldsymbol L_1 \times \boldsymbol L_2)} ,
\\
\boldsymbol r_{2,\pm} &= \frac{(1-\epsilon_2^2)a_2 \boldsymbol L_1 \times \boldsymbol L_2}{\pm|\boldsymbol L_1 \times \boldsymbol L_2| + \boldsymbol\epsilon_2\bullet(\boldsymbol L_1 \times \boldsymbol L_2)} 
.    
\end{align*}

\subsection{Pair of closest points between two orbits}
\label{Sec4.2}
Next, we approximate the points where the MOID is found. In order to do so, we consider the tangent lines of the orbits at the points $\boldsymbol r_1$ and $\boldsymbol r_2$ of intersection with the nodal line.
The tangent lines point in the direction of the velocities $\boldsymbol v_1$ and $\boldsymbol v_2$ at $\boldsymbol r_1$ and $\boldsymbol r_2$. These can be found using Eq.~(\ref{decomp}). The distance between the two lines is given by
\[
d = \bigg| (\boldsymbol r_2-\boldsymbol r_1)\bullet\frac{(\boldsymbol v_1\times\boldsymbol v_2)}{|\boldsymbol v_1\times\boldsymbol v_2|} \bigg|
.
\]
This is the projection of the difference vector onto the direction of shortest distance.
We refer to the two points on the lines where the distance is minimal as  $\boldsymbol r_1'$ and $\boldsymbol r_2'$. These positions
are given by
\begin{align*}
\boldsymbol r_1'
&= \boldsymbol r_1 + \bigg[ (\boldsymbol r_2-\boldsymbol r_1) \bullet \frac{|\boldsymbol v_2|^2\boldsymbol v_1 - (\boldsymbol v_1\bullet\boldsymbol v_2)\boldsymbol v_2}{|\boldsymbol v_1|^2|\boldsymbol v_2|^2-(\boldsymbol v_1\bullet\boldsymbol v_2)^2} \bigg] \, \boldsymbol v_1
, \\
&= \boldsymbol r_1 + \bigg[ (\boldsymbol r_2-\boldsymbol r_1) \bullet \frac{\boldsymbol v_2 \times (\boldsymbol v_1\times \boldsymbol v_2)}{|\boldsymbol v_1\times\boldsymbol v_2|^2} \bigg] \, \boldsymbol v_1
, \\
\boldsymbol r_2'
&= \boldsymbol r_2 + \bigg[ (\boldsymbol r_2-\boldsymbol r_1) \bullet \frac{\boldsymbol v_1 \times (\boldsymbol v_1\times \boldsymbol v_2)}{|\boldsymbol v_1\times\boldsymbol v_2|^2} \bigg] \, \boldsymbol v_2
.
\end{align*}
One may verify that $(\boldsymbol r'_2-\boldsymbol r'_1)\bullet\boldsymbol v_1=(\boldsymbol r'_2-\boldsymbol r'_1)\bullet\boldsymbol v_2=0$. This proves that the minimal distance between the lines is realized  at the points $\boldsymbol r_1'$ and $\boldsymbol r_2'$.
We also have that $(\boldsymbol r'_2-\boldsymbol r'_1)\bullet(\boldsymbol r_2-\boldsymbol r_1)=d^2$, implying that $|\boldsymbol r'_2-\boldsymbol r'_1|=d<|\boldsymbol r_2-\boldsymbol r_1|$.

If the inclination between the orbital planes, which is given by $I\approx K/L_1L_2$, is not much larger than $(s_1+s_2)/(a_1+a_2)$, the linear approximation is inaccurate. One can improve this approximation by reducing the lengths of $\boldsymbol r'_i$ , so that the points lie on the respective orbits, and then finding the shortest distance between the tangent lines to the orbits at these new points. Here, one may iterate as in the method of Newton Raphson.

\section{Calculating the earliest passage of the crossing}
\label{Sec5}
In this section, we derive expressions for the time it takes a particle on a Kepler orbit to go from $\vec r^0$ to $\vec r^1$. In the algorithm, $\vec r^0$ is the particle's creation point and $\vec r^1$ is the collision point. A standard approach is to use the eccentric anomaly $E$ values at these points. However, $E$ becomes ill defined for small eccentricities\footnote{The need for the following calculations was pointed out by \citet{Soliman2022}. 
}.
Here, we derive the exact expression, Eq.~(\ref{passage}), that does not depend on the value of $E$;  only the difference between the values of $E$ enters the derivation.

Kepler's second law states that the position vector sweeps out equal areas in equal times. The swept area can be decomposed into a segment of the ellipse (with an angle determined by the eccentric anomaly) plus the triangle between $-a\vec\epsilon$, $\vec 0,$ and $\vec r^0$ minus the triangle between $-a\vec\epsilon,$  $\vec 0,$ and $\vec r^1$ in Fig.~\ref{Figure11}(a). Therefore,
\[
\text{area} = \underbrace{\frac{(E^1-E^0)ab}{2}}_\text{ellipse segment} + \underbrace{\frac{a\vec\epsilon \times \vec r^0}{2} \bullet \hat{\vec L}}_\text{added triangle} -
\underbrace{\frac{a\vec\epsilon \times \vec r^1}{2} \bullet \hat{\vec L}}_\text{subtracted triangle}
.
\]
Kepler's second law therefore implies that the time it takes a body to move from $\vec r^0$ to $\vec r^1$ is equal to
\begin{equation}
t^1 - t^0 =
\frac{\text{area}}{\pi ab} T =
\frac{2\ \text{area}}{\omega ab} =
\frac{E^1-E^0}{\omega} - \frac{\vec\epsilon \times (\vec r^1-\vec r^0)}{\omega b} \bullet \hat{\vec L}
.
\label{deltat}
\end{equation}
As the ellipse has its major axis pointing in the direction of $\vec\epsilon$, we can transform the elliptic orbit into the circle in Fig.~\ref{Figure11}(b) with the linear transformation
\[
\vec r \xrightarrow{\text{squeeze}} \bigg( 1 - \frac{\vec\epsilon\ \vec \epsilon^\text{T}}{\epsilon^2} + \frac{b}{a} \frac{\vec\epsilon\ \vec \epsilon^\text{T}}{\epsilon^2} \bigg) \vec r = \vec r - \frac{a\vec \epsilon\bullet\vec r}{a+b} \vec\epsilon 
.
\]
This squeezes the semimajor axes by a factor $b/a$, and leaves the semiminor axes intact. The eccentric anomaly is defined with respect to the center of the circle. We therefore require the vector pointing from the center to the point on the squeezed $\vec r$. This vector is found by
\[
\vec r \xrightarrow{\text{translate}} \vec r + (a\vec \epsilon) \xrightarrow{\text{squeeze}} \bigg(\vec r - \frac{a\vec \epsilon\bullet\vec r}{a+b} \vec\epsilon \bigg) + (b\vec\epsilon)
.
\]
When $\vec r$ is on the orbit, the transformed vector has a length of $b$. The cosine of the difference in eccentric anomalies is the dot product between the directions of the squeezed vectors:
\[
\cos(E^1-E^0) =
\frac{1}{b^2} \Big( \vec r^1 - a\vec\epsilon\frac{\vec \epsilon\bullet \vec r^1}{a+b} + b\vec\epsilon \Big) \bullet \Big( \vec r^0 - a\vec\epsilon\frac{\vec \epsilon\bullet \vec r^0}{a+b} + b\vec\epsilon \Big)
.
\]
This simplifies to
\begin{equation}
\cos(E^1-E^0) =
\frac{\vec r^1\bullet\vec r^0}{b^2} + \frac{(\vec r^1+\vec r^0)\bullet\vec\epsilon}{a} - \frac{\vec r^1\bullet\vec\epsilon\ \vec\epsilon\bullet \vec r^0}{b^2}  + \epsilon^2
.
\label{cos}
\end{equation}
By noting that the cross product between the two direction vectors gives us the sine, we find in terms of the position vectors:
\[
\sin(E^1-E^0) =
\frac{\vec r^0\times\vec r^1}{ab}\bullet\hat{\vec L} + \frac{\vec\epsilon\times(\vec r^1-\vec r^0)}{b} \bullet \hat{\vec L}
.
\]
If we combine this with Eq.~(\ref{deltat}), we obtain Eq.~(2.69) in \citet{Murray2009} for the so-called $g$-function:
\[
t^1 - t^0 = \frac{E^1 - E^0 - \sin(E^1-E^0)}{\omega} + \underbrace{\frac{m(\vec r^0\times \vec r^1)\bullet\vec L}{L^2}}_{g\text{-function}}
.
\]
Although this Equation~is remarkable because it does not contain $\epsilon$ or the values $a$, $b$, we will not need it.

Differentiating Eq.~(\ref{cos}) with respect to time $t^1$ in the endpoint gives another equation:
\begin{equation}
\frac{-\omega a}{r^1}\sin(E^1-E^0) =
\frac{\vec v^1\bullet\vec r^0}{b^2} + \frac{\vec v^1\bullet\vec\epsilon}{a} - \frac{\vec v^1\bullet\vec\epsilon\ \vec\epsilon\bullet \vec r^0}{b^2}
.\label{sin}
\end{equation}
These results can be verified by direct substitution of Eqs.~(\ref{orbitvec}), (\ref{vvec}), and (\ref{epsvec}) into the right-hand side of Eq.~(12).
Equation~(\ref{deltat}) with the smallest non-negative value for $E^1-E^0$ that satisfies Eqs.~(\ref{cos}) and (\ref{sin}) gives the time to get from $\vec r^0$ to $\vec r^1$.

\section{Calculating the time to collision}
\label{Sec6}
We consider two planets~1 and 2, with a MOID of less than $s_1+s_2$, and we want to determine the time at which the planets collide. To this end, let $\boldsymbol r_1$ be the point on orbit~1 with minimal distance to orbit~2, and $\boldsymbol r_2$ the corresponding point on orbit~2 that is closest to $\boldsymbol r_1$. Let $\boldsymbol v_1$ and $\boldsymbol v_2$ be the respective velocities of the planets if they pass these points. Now, assume that there is a possible collision:
\[
|\boldsymbol r_2-\boldsymbol r_1| < s_1 +s_2
.
\]
Let $t^1_1$ be the first time that planet~1 passes $\boldsymbol r_1$, and $t^1_2$ be the first time planet~2 passes $\boldsymbol r_2$. A collision occurs at time $t(k,l)$ when both planets are near the points where the distance to the other ellipse is minimal. At that time, planet~1 then passes the point for the $k$-th time, and planet~2 for the $l$-th time. The collision is therefore at
\[
t(k,l) = t^1_1 + T_1 k + \mathrm dt_1 = t^1_2 + T_2 l + \mathrm dt_2
.
\]
Here $k$ and $l$ are integers and $\mathrm dt_1$ and $\mathrm dt_2$ are small shifts that allow for the fact that the planets only need to be close to the point where the distance between the orbits is minimal. Because the algorithm moves forward in time,
\[
k \geq 0 , \quad l \geq 0
.
\]
The shifts in time from the point of closest approach are therefore
\begin{equation}
\mathrm dt_1 = t(k,l) - t^1_1 - T_1 k , \quad
\mathrm dt_2 = t(k,l) - t^1_2 - T_2 l
,
\label{defkl}
\end{equation}
and these need to be small. We linearize the motion about the collision time $t(k,l)$, as
\[
\boldsymbol r_1 +\boldsymbol v_1 \mathrm dt_1 , \quad \boldsymbol r_2 +\boldsymbol v_2 \mathrm dt_2
.
\]
By solving for the closest approach between the two particles (in contrast to the MOID, the smallness of the differentials will be a consequence of the fact that the minimal distance
for colliding particles is smaller than $s_1+s_2$. For this, we introduce the difference vectors
\[
\boldsymbol r_{12}(t) = \boldsymbol r_2 + \boldsymbol v_2 \mathrm dt_2 - \boldsymbol r_1 - \boldsymbol v_1 \mathrm dt_1 , \quad
\boldsymbol v_{12} = \boldsymbol v_2 - \boldsymbol v_1
.
\]
We note that $\mathrm dt_1$ and $\mathrm dt_2$ need to be considered as functions of the collision time $t$. At this time $t$, the distance is minimal, which 
is at (see Eq.~(\ref{timemindist}) in Appendix~\ref{AppA})
\[
t(k,l) = \frac{-\boldsymbol v_{12} \bullet \boldsymbol r_{12}(0)}{v_{12}^2}
,
\]
with $v_{12}=|\boldsymbol v_{12}|$. The value of the distance must be smaller than the sum of the planet radii \citep[see][]{JeongAhn2017}:
\[
\frac{|\boldsymbol v_{12}\times \boldsymbol r_{12}(0)|}{v_{12}} < s_1 + s_2
.
\]
When we expand this equation, we find
\[
\Big| \Big( \boldsymbol r_2 - \boldsymbol r_1 - \boldsymbol v_2 t^1_2 - \boldsymbol v_2 T_2 l + \boldsymbol v_1 t^1_1 + \boldsymbol v_1 T_1 k \Big) \times \boldsymbol v_{12}\Big| < (s_1+s_2) v_{12}
.
\]
Because $(\boldsymbol r_2-\boldsymbol r_1)\bullet \boldsymbol v_1=(\boldsymbol r_2-\boldsymbol r_1)\bullet \boldsymbol v_2=0$, this is equivalent to
\[
\Big| \Big( \boldsymbol v_1 t^1_1 + \boldsymbol v_1 T_1 k - \boldsymbol v_2 t^1_2 - \boldsymbol v_2 T_2 l \Big) \times \boldsymbol v_{12}\Big| < \!\sqrt{(s_1+s_2)^2 - |\boldsymbol r_2 - \boldsymbol r_1|^2} v_{12}
.
\]
Using $\boldsymbol v_{12}=\boldsymbol v_2-\boldsymbol v_1$, this can be further simplified to
\[
\Big| t^1_1 + T_1 k - t^1_2 - T_2 l \Big|  < \frac{\sqrt{(s_1+s_2)^2 - |\boldsymbol r_2 - \boldsymbol r_1|^2} v_{12}}{|\boldsymbol v_1\times\boldsymbol v_2|}
.
\]
We need to find the smallest non-negative integers $k$, $l$ for which this inequality is satisfied. We can now recast the problem of the time to collision as finding the smallest integers $k$, $l,$ so that
\begin{equation}
1 - \delta < kp - lq < 1 + \delta
\label{irrational}
.
\end{equation}
In this inequality, we use dimensionless parameters $p$, $q$, $\delta$, which are defined as
\[
p = \frac{T_1}{|t^1_1-t^1_2|} , \quad
q = \frac{T_2}{|t^1_1-t^1_2|} , \quad
\delta = \frac{\sqrt{(s_1+s_2)^2 - |\boldsymbol r_2 - \boldsymbol r_1|^2} v_{12}}{|t^1_1-t^1_2| |\boldsymbol v_1\times\boldsymbol v_2|}
.
\]
The linearization of the motion around the crossing points of the nodal line translates the collision problem into integer linear programming (in two dimensions).
We can find the exact solution in a few steps, even if $k$ and $l$ turn out to be very large numbers.

\begin{figure}
\centering
\includegraphics{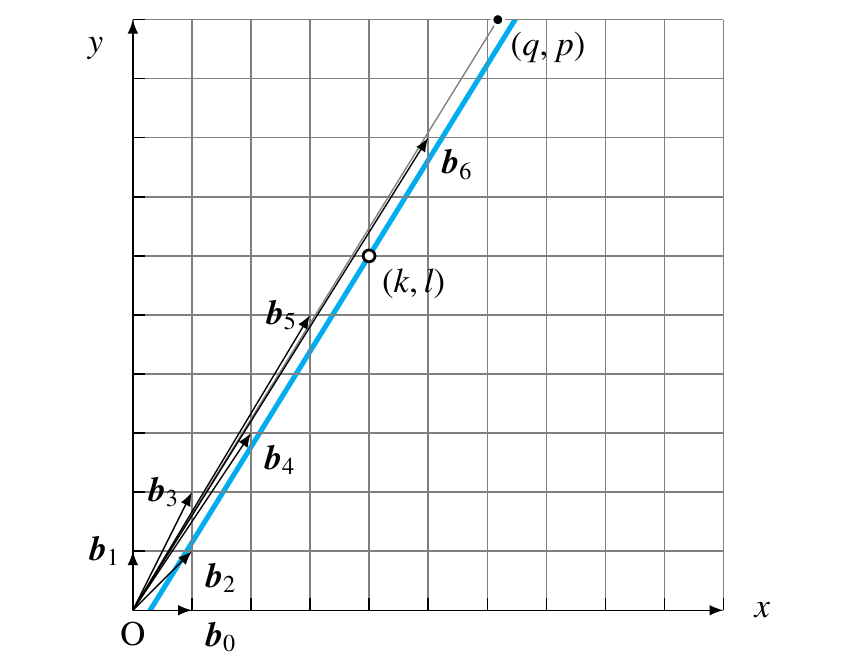}
\caption{Search space. The problem of finding the collision time for two planets
is equivalent to finding the grid point $(k,l)$ in the small \emph{blue} band that is closest to the origin. The horizontal and vertical axes are given by $x=(t-t_1^1)/T_1$ and $y=(t-t_2^1)/T_2$ in Eq.~(\ref{defkl}), respectively. The slope is the ratio $p/q=T_1/T_2$ of orbital periods. The center of the intersection with the horizontal axis is $T_2/|t_1^1-t_2^1|$, where $|t_1^1-t_2^1|$ is the difference between the time-of-passage of the collision point for the two planets. The width of the band depends on the minimal distance of the orbits and the planet radii. The solution can be rapidly found using the bases $\big\{\boldsymbol b_{2n},\boldsymbol b_{2n+1}\big\}$.
}
\label{Figure12}
\end{figure}

We assume, without loss of generality, that $T_1>T_2$. Consequently,
$p>q>0$ and $p>1$. Equation~(\ref{irrational}) says that we need an integer linear combination of the irrationals $p$ and $q$ that is within a distance $\delta$ from $1$. We therefore need to find the point in the grid $\mathbb N^2$ closest to the origin that lies in between the lines
\[
x = \bigg(\frac{q}{p}\bigg) y + \frac{1-\delta}{p}  \quad \text{and} \quad
x = \bigg(\frac{q}{p}\bigg) y + \frac{1+\delta}{p}
\]
in the first quadrant of the $xy$-plane ($\mathbb R^2$). This is shown in Fig.~\ref{Figure12}. The (horizontal) width of the narrow band is $\delta/p,$ which will be of order of magnitude of
$s/a$. For the following solution method to give the correct values of $k$ and $l$, the numerical accuracy needs to be smaller than this number.

\begin{figure}
\centering
\includegraphics{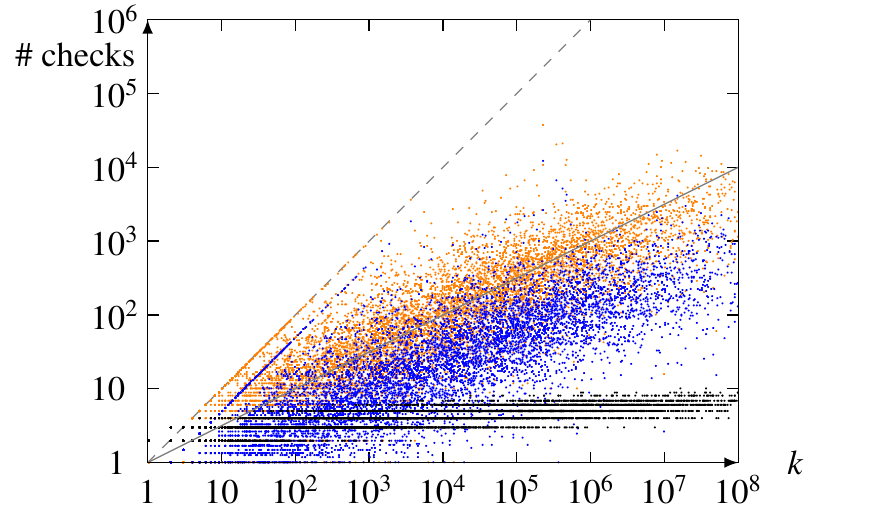}
\caption{Numerical checks needed to calculate the exact collision time for random collision pairs.
Here we show the number of checks (\emph{orange}), the number of iterations (\emph{black}), and the average number of checks per iteration (\emph{blue}) performed by the algorithm. On the horizontal axis is the number of orbits $k$ before the collision, which is
the solution calculated by the iteration scheme. The \emph{gray} line is $\#\text{checks}=k^{1/2}$ and shows the trend.}
\label{Figure13}
\end{figure}

\subsection{Deterministic collision time}
\label{Sec5.1}
Here we present an iteration scheme that finds $(k,l)$ in $\log(q/\delta)$ steps. We will need the continued-fraction representation of $p/q$. This is found from the recursion relation that calculates the successive remainders:
\[
q_0 = p , \quad q_1 = q , \quad
q_{n+2} = (q_n \;\text{mod}\; q_{n+1})
, \quad \textrm{for} \quad n = 0,1,2,\ldots
\]
This is similar to the Euclidian algorithm for finding the greatest common divisor. For rational $p/q$, the sequence becomes zero in finite steps, and for irrational $p/q$ the sequence decreases to zero \citep{Khinchin1964,Rockett1992}:
\[
0 < q_{n+1} < q_n \longrightarrow 0 , \quad \textrm{as} \quad n \longrightarrow \infty
.
\]
The $q_n$ are all integer linear combinations of $p$, $q$, and are therefore elements of $\mathbb Z\text{-span}\big\{p,q\big\}$. When we write $q_n=k_n p-l_n q$, then the rationals $l_n/k_n$ are precisely the successive convergents of the continued-fraction representation.

Next, we define the bases $\big\{\boldsymbol b_n, \boldsymbol b_{n+1}\big\}$ for $\mathbb Z^2$, with slopes equal to the fractions:
\[
\boldsymbol b_n = (-1)^n \bigg(\begin{matrix} k_n \\ l_n \end{matrix}\bigg)
, \quad
\frac{l_{2n}}{k_{2n}} < \frac{p}{q} < \frac{l_{2n+1}}{k_{2n+1}}
.
\]
The slopes of the even base vectors increase and the slopes of the odd base vectors decrease to $p/q$. This is shown in Fig.~\ref{Figure12}. The sequence of remainders may be found from
\[
\bigg(\begin{matrix} q_{n+2} \\ q_{n+1} \end{matrix}\bigg) =
\bigg(\begin{matrix} -\lfloor q_n/q_{n+1}\rfloor & 1 \\ 1 & 0 \end{matrix}\bigg)
\bigg(\begin{matrix} q_{n+1} \\ q_n \end{matrix}\bigg)
,
\]
and base vectors can be found from the recursion relation:
\[
\boldsymbol b_0 = \bigg(\begin{matrix} 1 \\ 0 \end{matrix}\bigg) , \quad
\boldsymbol b_1 = \bigg(\begin{matrix} 0 \\ 1 \end{matrix}\bigg) , \quad
\boldsymbol b_{n+2} = \boldsymbol b_n + \biggl\lfloor \frac{q_n}{q_{n+1}} \biggr\rfloor \boldsymbol b_{n+1}
.
\]
Therefore, each basis is related to the preceding basis by the transformation
\[
\bigg(\begin{matrix} \boldsymbol b_{n+1} \bigg| & \boldsymbol b_{n+2} \end{matrix}\bigg) =
\bigg(\begin{matrix} \boldsymbol b_n \bigg| & \boldsymbol b_{n+1} \end{matrix}\bigg)
\bigg(\begin{matrix} 0 & 1 \\ 1 & \lfloor q_n/q_{n+1}\rfloor \end{matrix}\bigg)
.
\]
The transformation matrix is unimodular, which implies that the transformation is a bijection between the points of $\mathbb Z^2$. The integer coordinates $(x_n,y_n)$ in each base are defined by
\[
\bigg(\begin{matrix} x \\ y \end{matrix}\bigg) = x_n \boldsymbol b_n + y_n \boldsymbol b_{n+1}
,
\]
which means that $x_0=x$ and $y_0=y$. Substitution in Eq.~(\ref{irrational}) gives us the inequalities 
\[
1 - \delta < x_0 p - y_0 q = (-1)^n(x_n q_n - y_n q_{n+1}) < 1 + \delta
.
\]

Consider the following bases, which are composed of the successive even- and odd-numbered vectors:
\[
\big\{ \boldsymbol b_0 , \boldsymbol b_2 \big\} , \quad
\big\{ \boldsymbol b_1 , \boldsymbol b_3 \big\} , \quad
\big\{ \boldsymbol b_2 , \boldsymbol b_4 \big\} , \quad
\big\{ \boldsymbol b_3 , \boldsymbol b_5 \big\} , \ldots
\]
The positive spans (linear combinations) of these pairs partition the first quadrant of $\mathbb R^2$ into segments (see again Fig.~\ref{Figure12}). As we may assume that $(k,l)=(0,0)$ is not a solution, the band intersects the $y$-axis at negative $y$ and therefore lies below the segments spanned by the odd bases. However, in any even basis, the intersection of the band with a segment is a trapezoid.
Because the union of all segments is the entire first quadrant, the solution of Eq.~(\ref{irrational}) must lie in one of the trapezoids. However, the pairs of even and odd base vectors do not form a unimodular matrix, and therefore they do not necessarily span $\mathbb Z^2$. For this reason, we must instead search in the original bases $\big\{\boldsymbol b_n,\boldsymbol b_{n+1}\big\}$ inside the parallelogram of the strip between $y'=0$ and the $y'$-value where the bottom of the strip intersects $\text{span}\big\{\boldsymbol b_{n+2}\big\}$. This is the top-right point and the bottom-right point of the subsequent parallelogram. The parallelogram has corner points:
\begin{align*}
(x',y') =& \Big(\tfrac{1-\delta}{q_n},0\Big) , \quad
\Big(\tfrac{1-\delta}{q_n}+\tfrac{(1+\delta)(q_n-q_{n+2})}{q_n q_{n+2}},\tfrac{(1+\delta)(q_n-q_{n+2})}{q_{n+1}q_{n+2}}\Big) , \\
&
\Big(\tfrac{1+\delta}{q_n},0\Big) , \quad
 \Big( \tfrac{1+\delta}{q_{n+2}},\tfrac{(1+\delta)(q_n-q_{n+2})}{q_{n+1}q_{n+2}}\Big)
,
\end{align*}
where $n$ is even and the coordinates are defined by
\[
\bigg(\begin{matrix} x \\ y \end{matrix}\bigg) =
x' \boldsymbol b_n + y' \boldsymbol b_{n+1} = 
\bigg(\begin{matrix} x' k_n - y' k_{n+1} \\ x' l_n - y' l_{n+1} \end{matrix}\bigg) 
.
\]
Successive parallelograms overlap. One needs to search one entire parallelogram first before going to the next, because the integer point closest to the origin will be found at the earliest occasion. Also, for each $y'$-value, only the lowest integer $x'$-value to the right of the left line segment needs to be checked. \footnote{It was pointed out by \citet{Schouten2022} that by looping through the integer $x$ values instead of the integer $y$ values the number of checks is significantly lower.}

When the calculated orbital periods $T_1$ and $T_2$ have a ratio close to that of two small integers, the planets could be in mean-motion resonance and may never collide (as is the case with Neptune and Pluto). The method neglects these cases, and will erroneously find a collision at a high $k$ number. If the continued fraction is actually finite (because $p/q$ is rational), the strip has an integer slope at the final step.
Figure~\ref{Figure13} shows the results of a numerical test with 
random collision pairs. We find that the total number of checks grows as $k^{1/2}$, while the number of iterations grows as $\log k$ with the solution $k$.

\subsection{Stochastic collision times}
\label{Sec5.2}
For small $\delta$ and irrational $p/q$, a generic solution $(k,l)$ will form a pair of large integers, with
\[
\frac{l}{k} \approx \frac{p}{q}
.
\]
The precise value depends very sensitively on $q$ and $p$. If we assume that we cannot obtain the required numerical accuracy to find the exact solution, we may use a statistical approach. The integer points $(k,l)$ are uniformly distributed over the plane. If we assume that the points are statistically independent (this it clearly an approximation valid for $\delta\ll 1$), the distribution is that of a Poisson point process. The probability that there is a grid point $(k,l)$ between the lines with $k$ in the interval $[x,x+\mathrm dx]$ is equal to the area of the small parallelogram. This area is $(2\delta/q)\mathrm dx$. Now, the area between the lines below $k=x$ is given by
\[
\frac{2\delta}{q}x - \frac{2\delta-4\delta^2}{pq} \approx \frac{2\delta}{q}x , \quad\text{for} \quad x \gg 1
.
\]
Therefore, the respective probabilities for the solution $k$ to be found above $x$ and below $x$ are
\begin{align*}
\mathrm{Prob}(k>x) &= \lim_{\mathrm dx\longrightarrow 0} \Big( 1 - \tfrac{2\delta}{q}\mathrm dx \Big)^{x/\mathrm dx} = \mathrm e^{-2x\delta/q}
, \\
\mathrm{Prob}(k\leq x) &= 1 - \mathrm e^{-2x\delta/q}
.
\end{align*}
The latter formula is the cumulative distribution function for $k$. We obtain a realization by generating a random real number $\xi$ inside the interval $[0,1]$ and use
\[
k = \frac{(-\log\xi)q}{2\delta} , \quad l = \frac{(-\log\xi)p}{2\delta} - \frac{1}{q}
.
\]
Because these values are large, rounding off to the nearest integer is not important. The time of the collision is then given by
\[
t = t_1^1 + kT^1 = t_1^1 + \frac{(-\log\xi)T_1T_2 |\boldsymbol v_1\times\boldsymbol v_2|}{2\sqrt{(s_1+s_2)^2 - |\boldsymbol r_2 - \boldsymbol r_1|^2} v_{12}}
.
\]
The formula for the average waiting time 
$\overline kT^1$ is precisely the reciprocal of the collision probability for one orbit. The fact that this reproduces the formula Eq.~(23) in \citet{Opik1951}, Eq.~(29) in \citet{JeongAhn2017}, and Eqs.~(2-3) in \citet{Diserens2020} for this probability $P$ validates our Eq.~(\ref{irrational}).

The Monte Carlo method for finding the collision time also follows from assuming homogeneous distributions of the mean anomalies of two fixed Kepler orbits, as in \"Opik's scheme. This method assumes large $k$, implying that the initial crossing time $t_1$ cannot be precisely known (due to numerical inaccuracy or neglected physics effects). However, when the system contains one or more large planets, the collision or nearby passage could happen after a few revolutions, that is,\ for small $k$.

\subsection{Including orbital precession}
\label{Sec6.3}
\begin{figure}[t]
\centering
\includegraphics{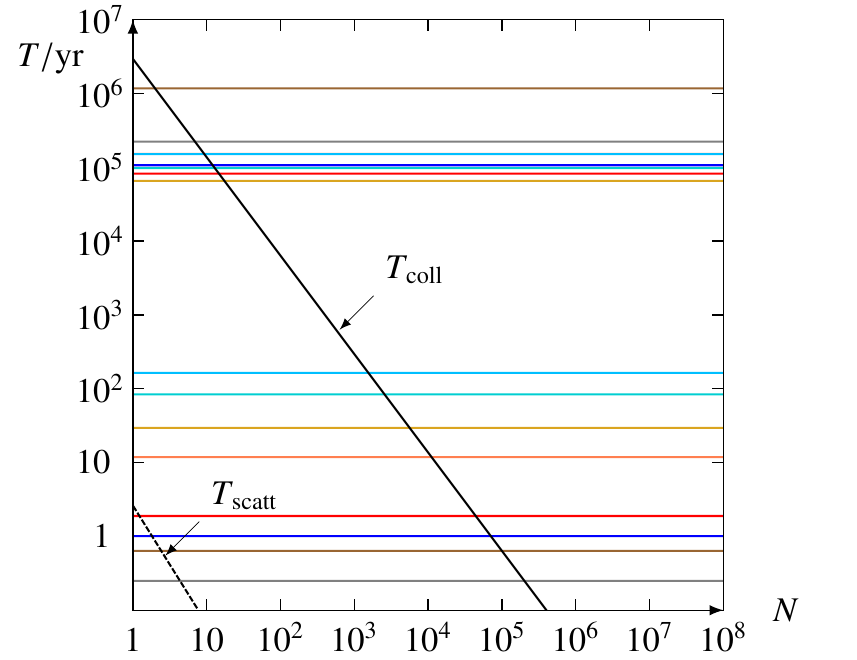}
\caption{Timescale vs.\ particle count $N$. The horizontal lines are the orbital periods $T_j$ and the precession periods $T_{\text{prec}j}$ of the Solar System planets (\emph{gray}, \emph{brown}, \emph{blue}, \emph{red}, \emph{coral}, \emph{golden}, \emph{turquoise}, \emph{azure} are Mercury through Neptune) from \citet{Murray2009}. The \emph{black} curve shows the collision time $T_\text{coll}$ estimated for a disk with $a=4\text{au}$, $I=.1$, and a mass of three Earths containing $N$ particles of Earth density. For $N<10$, the collision time is comparable to the precession time ($10^5\text{yr}$). The radius of gravitational influence is roughly $10^3$ times the planet radius $s$; this determines $T_\text{scatt}$, the \emph{dashed} curve.}
\label{Figure14}
\end{figure}

Our method for calculating the collision time outlined in the previous subsection assumes perfect Kepler orbits. However, if one intends to make accurate predictions over long timescales, the slow precession of the periapsis and of the orbital plane cannot be neglected. For example, the perihelion shift for the planet Jupiter in one revolution is about twice the planet's diameter (see Fig.~\ref{Figure14} for the precession rates for the Solar System planets). For a planet ring system, precession is mainly due to the oblateness of the planet. Hence, if one is not just interested in statistical averages,
then the secular dynamics must be included on the orbital timescale.

A method to remedy this problem is to numerically propagate the Laplace-Lagrange equations.
The time-steps are now set by the secular timescale $\mathrm dt \lessapprox T_\text{prec}$. There are $N^2$ terms in the system of differential equations and there are $N^2\epsilon/2$ collision possibilities, which need to be evaluated. The theoretical algorithmic efficiency is shown by the orange curve in Fig.~\ref{Figure8}.

It may be possible to include the collision detection in the following way. At each time-step (now shorter than the collision time), one calculates the instantaneous change is the orbital elements $\epsilon$, $\varpi$, $I$, $\ascnode$.
One expresses the resulting linear change in the points $\boldsymbol r_j$ and velocities $\boldsymbol v_j$ near the points of closest approach, as linear functions in the passage numbers $k$ and $l$. The step where the minimal distance $d$ is compared with the sum $s_i+s_j$ is skipped. Instead one directly uses the modified inequality Eq.~(\ref{irrational}).  This would then lead to a problem from integer linear programming, as before. For this modified case, we would expect the two lines bounding our search domain to be nonparallel. The solution for the collision problem has a spatial accuracy of less than a planet radius on the longer timescale where the perihelion shift can be approximated as linear motion. A thorough development of this idea is a possible direction for follow-up research.

\section{Conclusions}
\label{Sec7}
We describe an algorithm that simulates collisional Keplerian systems: $N$ bodies in the Coulomb potential of a central mass. The method uses the orbital elements and has three basis ingredients, of which the third is novel:
\begin{inparaenum}[(i)]
\item
For a new particle $i$, a small set of possible collision candidates $j$ is selected using the apoapsis/periapsis filter.
\item The MOIDs between the particle pairs $(i,j)$ can be approximated numerically.
\item For the pairs $(i,j)$ on a collision trajectory, one can obtain the collision time $t_{(i,j)}$ with integer linear programming.
\end{inparaenum}
During the simulation, sorted lists of the particles and the collision pairs are maintained. The algorithm steps from one collision to the next as it updates the particle orbits and collision possibilities.

We show that the problem of finding the collision time is mathematically equivalent to the problem in integer linear programming of finding the grid point $(k,l)$ in $\mathbb N^2$ between two parallel lines that is closest to the origin. The exact solution uses the continued-fraction representation of the ratio $T_i/T_j$ of the orbital periods.

Because at most $N$ new collision possibilities have to be added to the list, less than $N$ interactions need to be considered at each step. The length of the collision list is $O(N^2s/a)$ and the total number of collisions is $O(N^2s^2/a^2)$, resulting in an algorithmic efficiency of $O(N^4s^3/a^3)$. This may be compared to the efficiency $O(N^{4/3}\log N)$ of a numerical integration propagator with collision detection (tree code or spatial hashing), which is independent of the particle radius $s$. In the astronomical applications, the radii are usually small compared to the orbits. The collisions are therefore rare, and the proposed collision-detection method can be fast. However, the perturbations we neglect become increasingly important, and, at the same time, the result becomes progressively sensitive to the initial state. Needless to say, including collisions is important, even if they are rare. Studying statistics of outcomes of the dynamics requires many simulations with near-identical initial states. For relatively small particle numbers, say for $N<10^6$, the individual realizations can be fast.

\begin{acknowledgements}
The author would like to thank John Chambers for acting as referee and
for improving the algorithm, and Dylan Aliberti, Aron Schouten, and Philip Soliman for their meticulous checking and verification of the formulas and algorithms in this paper.
\end{acknowledgements}

\bibliographystyle{aa}
\bibliography{references}{}

\begin{appendix}
\section{Closest approach}
\label{AppA}
Consider two non-interacting particles in linear motion:
\[
\boldsymbol r_1(t) = \boldsymbol r_1 + \boldsymbol v_1 t , \quad
\boldsymbol r_2(t) = \boldsymbol r_2 + \boldsymbol v_2 t . \quad
\]
We call the intitial distance vector and the relative velocity:
\[
\boldsymbol d = \boldsymbol r_2 - \boldsymbol r_1 , \quad
\boldsymbol u = \boldsymbol v_2 - \boldsymbol v_1
.
\]
We want to decide if there is a collision in the interval $[0,\mathrm dt]$. Therefore, we calculate the distance between the particles at $t=0$ and at $t=\mathrm dt$ to see if there is a collision at the endpoints:
\[
|\boldsymbol d| < s_1 + s_2 , \quad
|\boldsymbol d + \boldsymbol u\, \mathrm dt| < s_1 + s_2
.
\]
If not, the only possibility for a collision on the interval is that the distance obtains a minimum on the (interior) of the interval.
This means that the relative velocity in the direction between the particles is first decreasing and then increasing:
\[
\boldsymbol u\bullet \boldsymbol d < 0 , \quad
\boldsymbol u\bullet (\boldsymbol d + \boldsymbol u\, \mathrm dt) > 0
.
\]
The time of minimal distance is at
\begin{equation}
t = \frac{- \boldsymbol u\bullet\boldsymbol d}{u^2}
\label{timemindist}
,
\end{equation}
which is then indeed between $0$ and $\mathrm dt$. We then decide if the distance at this time $t$ is smaller than the sum of the radii. When we substitute $t$ back into the Equation for the distance, we find that it is equal to the component of $\boldsymbol d$  perpendicular to the direction of $\boldsymbol u$. Hence, the condition for a collision is equivalent to:
\[
\frac{|\boldsymbol u \times \boldsymbol d|}{|\boldsymbol u|} < s_1+s_2
.
\]

\section{Orbital elements}
\label{AppB}
Consider a single particle of mass $m$ in a Kepler orbit about the central mass $M$. The orbit is an ellipse in a fixed plane. The angular momentum vector is defined by
\[
\boldsymbol L = \boldsymbol r \times m\boldsymbol v
.
\]
The \emph{Laplace-Runge-Lenz vector} is proportional to the dimensionless eccentricity vector:
\[
\boldsymbol\epsilon = \frac{\bf{LRL}}{GMm^2} = \frac{\boldsymbol v\times \boldsymbol L}{GMm} - \frac{\boldsymbol r}{r}
.
\]
The orbit is fixed by the orthogonal pair of vectors $\boldsymbol L$ and $\boldsymbol\epsilon$.

For the problem of solving the MOID, we need a formula that expresses the velocity $\boldsymbol v$ along the orbit as a function of the position $\boldsymbol r$ and the orbital elements. Given $\boldsymbol L$, $\boldsymbol\epsilon$, $\boldsymbol r,$ we equate
\[
\boldsymbol L \times \bigg( \boldsymbol\epsilon + \frac{\boldsymbol r}{r}\bigg) =
\frac{\boldsymbol L\times (\boldsymbol v\times\boldsymbol L)}{GMm} =
\frac{L^2 \boldsymbol v}{GMm}
,
\]
and therefore the velocity can be expressed as:
\begin{equation}
\boldsymbol v =
\frac{GMm}{L^2} \boldsymbol L \times \bigg( \boldsymbol\epsilon + \frac{\boldsymbol r}{r}\bigg)
.
\label{decomp}
\end{equation}
The Equation~for the energy is called the \emph{vis-viva equation}
\[
\frac{\mathrm{Energy}}{m} = \frac{v^2}{2} - \frac{GM}{r} = -\frac{GM}{2a}
.
\]
With this Equation~and \emph{Kepler's third law},
\[
\omega^2a^3 = GM
,
\]
the orbital elements $a$, $\epsilon$ and mean motion $\omega$ can be calculated from position and velocity:
\[
a = \frac{1}{1-\epsilon^2} \frac{L^2}{GMm^2} = \frac{1}{\displaystyle\frac{2}{r}-\frac{v^2}{GM}}
, \quad
\omega = \sqrt{\displaystyle\frac{GM}{a^3}}
.
\]

The orbit is parametrized by the true anomaly $\nu$ or the eccentric anomaly $E$. If we know the eccentric anomaly, we can calculate the time since periapsis from the \emph{Kepler equation}
\begin{equation}
t = \frac{E-\epsilon\sin E}{\omega}
.
\label{kepler}
\end{equation}
The formula for the radial distance in terms of the parameters is
\begin{equation}
r = \frac{b^2}{a+c\cos\nu} =
a - c\cos E
.
\label{orbit}
\end{equation}
The semimajor axes, $a$ and $b$, the distance from the center to a focus, $c$, and eccentricity $\epsilon$ are related by
\[
b = \sqrt{1-\epsilon^2} a , \quad c =\epsilon a , \quad a^2 = b^2 + c^2
.
\]
The position vector and the velocity vector can now be expressed as:
\begin{equation}
\boldsymbol r = \begin{pmatrix} x \\ y \\ z \end{pmatrix} =
r \mathscr{R} \begin{pmatrix} \cos\nu \\ \sin\nu \\ 0 \end{pmatrix} =
\mathscr{R} \begin{pmatrix} a \cos E - c \\ b\sin E \\ 0 \end{pmatrix}
,
\label{orbitvec}
\end{equation}
and, using Eqs.~(\ref{kepler}) and (\ref{orbit}),
\begin{equation}
\boldsymbol v = \dot{\boldsymbol r} =
\frac{\omega a}{r} \mathscr{R} \begin{pmatrix} -a\sin E \\ b\cos E \\ 0 \end{pmatrix} =
\frac{\omega a}{b} \mathscr{R} \begin{pmatrix} -a\sin\nu \\ a\cos \nu + c \\ 0 \end{pmatrix}
.
\label{vvec}
\end{equation}
In the expressions Eqs.~(\ref{orbitvec}) and (\ref{vvec}), $\mathscr{R}$ is a constant rotation matrix, which can be expressed as a product of three elementary rotations
\[
\mathscr{R} =
\begin{pmatrix} \cos\ascnode & -\sin\ascnode & 0 \\ \sin\ascnode & \cos\ascnode & 0 \\ 0 & 0 & 1 \end{pmatrix}
\begin{pmatrix} 1 & 0 & 0 \\ 0 & \cos I & -\sin I \\ 0 & \sin I & \cos I \end{pmatrix}
\begin{pmatrix} \cos\varpi & -\sin\varpi & 0 \\ \sin\varpi & \cos\varpi & 0 \\ 0 & 0 & 1 \end{pmatrix}
.
\]
Here, $\varpi$ is the argument of perihelion, $I$ is inclination, and $\ascnode$ is the ascending node. We have for the angular momentum and the eccenticity vectors,
\begin{equation}
\boldsymbol L =
L \mathscr{R} \begin{pmatrix} 0 \\ 0 \\ 1 \end{pmatrix} =
L \begin{pmatrix} \sin\ascnode\sin I \\ -\cos\ascnode\sin I \\ \cos I \end{pmatrix}
, \quad
L = m\omega ab
,
\label{Lvec}
\end{equation}
and
\begin{equation}
\boldsymbol\epsilon =
\epsilon \mathscr{R} \begin{pmatrix} 1 \\ 0 \\ 0 \end{pmatrix} =
\epsilon \begin{pmatrix} \cos\varpi\cos\ascnode - \sin\varpi\sin\ascnode\cos I \\ \cos\varpi\sin\ascnode + \sin\varpi\cos\ascnode\cos I \\ \sin\varpi\sin I \end{pmatrix}
, \quad
\epsilon = \frac{c}{a}
.
\label{epsvec}
\end{equation}
These five orbital elements $a$, $\epsilon$, $\varpi$, $\ascnode$, $I$ represent five independent conserved quantities. They can be found from the position and velocity vectors as follows. The equations at the beginning of this section give us $\boldsymbol L$ and $\boldsymbol \epsilon$ in terms of $\boldsymbol r$, $\boldsymbol v$. Then, the component $l=L_3$ can give us the angle $I$. The component $L_1$ can give us $\ascnode$, and $\epsilon_3$ can give us the angle $\varpi$.

\section{Elastic collisions}
\label{AppC}
When planets approach each other so closely that the gravity they exert on each other is of the same strength as the gravity from the star, the mutual gravitational interaction cannot be neglected. The close encounter can be approximated as a scattering event. Here, for the short duration of the encounter,  we neglect  the gravity from the central star (and of all other particles in the system). As no energy is dissipated, the process is an elastic collision. It is described by a hyperbolic orbit for the relative coordinate $\boldsymbol r_2-\boldsymbol r_1$. The hyperbola has parameters $a$, $b$, $c$, and is given  by the following equations in Cartesian and in polar coordinates
\[
\frac{(x-c)^2}{a^2} - \frac{y^2}{b^2} = 1
, \quad x \leq c-a , \quad
r = \frac{b^2}{a+c\cos\nu}
.
\]
The parameter $a$ is the semi-transverse axis of the hyperbola. It is a characteristic distance where the gravity between the two bodies becomes noticeable given the relative velocity. The semi-conjugate axis $b$ of the hyperbolic orbit is equal to the impact parameter. The semi-focal separation is $c$ (see Fig.~\ref{Figure9}). We have
\[
a = \frac{G(m_1+m_2)}{u^2}
, \quad
b = d = |\boldsymbol d| , \quad
c^2 = a^2 + b^2
.
\]
When the scattering between two planets is significant, the velocities of the two planets will change. The center-of-mass moves with velocity
\[
\boldsymbol v = \frac{m_1\boldsymbol v_1+m_2\boldsymbol v_2}{m_1+m_2}
.
\]
In the center of mass frame, both velocities rotate over an angle $\pi-2\arctan(b/a)$. This is described by:
\begin{align*}
\boldsymbol u' & = \frac{b^2-a^2}{c^2}\boldsymbol u - \frac{2au}{c^2} \boldsymbol d
, \\
\boldsymbol v'_1 &= \boldsymbol v - \frac{m_2}{m_1+m_2} \boldsymbol u'
, \\
\boldsymbol v'_2 &= \boldsymbol v + \frac{m_1}{m_1+m_2}\boldsymbol u'
.
\end{align*}

\section{Mathematica code}
The following Mathematica code tests the deterministic collision-time algorithm. It first finds the solution by considering all integers $k$, and then uses the continued-fraction representation. Both procedures give the same result for random orbital periods.
\label{AppD}
\begin{lstlisting}
tmax = 100;
T1 = RandomReal[tmax];
T2 = RandomReal[tmax];
If[T1 < T2, T1 = tmax - T1; T2 = tmax - T2];
t1 = RandomReal[T1];
t2 = RandomReal[T2];
deltat = Abs[t1 - t2];
p = T1/deltat
q = T2/deltat
del = deltat/10^5

qlist = {p, q};
n = 1;
While[qlist[[n + 1]] > del,
 qnm = qlist[[n]];
 n++;
 qn = qlist[[n]];
 AppendTo[qlist, qnm - Floor[qnm/qn] qn]
 ]
qlist

For[k = 1, k < 10^6, k++,
 l = Ceiling[p k/q - (1 + del)/q];
 If[l < p k/q - (1 - del)/q, Break[]]
 ]
Plot[
 {p x/q - (1 - del)/q, p x/q - (1 + del)/q},
 {x, k - del, k + del},
 PlotRange -> {l - del, l + del},
 GridLines -> {{k}, {l}},
 AspectRatio -> 1, Axes -> None,
 Frame -> None
 ]
k
l

q0 = p; q1 = q;
k0 = 1; k1 = 0;
l0 = 0; l1 = -1;
plot = {};
found = False;
For[n = 0, Not[found], n++,
 a0 = Floor[q0/q1];
 q2 = q0 - a0 q1;
 If[q2 == 0, Break[]];
 k2 = k0 - a0 k1;
 l2 = l0 - a0 l1;
 a1 = Floor[q1/q2];
 For[x = Ceiling[(1 - del)/q0],
  x < (1 + del)/q2, x++,
  y = Max[0, Ceiling[(q0 x - 1 - del)/q1]];
  If[1 - del < x q0 - y q1,
   k = x k0 - y k1;
   l = x l0 - y l1;
   found = True;
   Break[]
   ]
  ];
 q3 = q1 - a1 q2;
 If[q3 == 0, Break[]];
 k3 = k1 - a1 k2;
 l3 = l1 - a1 l2;
 q0 = q2; q1 = q3;
 k0 = k2; k1 = k3;
 l0 = l2; l1 = l3;
 ]
k
l

Plot[
 {p x/q - (1 - del)/q, p x/q - (1 + del)/q},
 {x, k - del, k + del},
 PlotRange -> {l - del, l + del},
 GridLines -> {{k}, {l}}, 
 AspectRatio -> 1, Axes -> None,
 Frame -> None
 ]
\end{lstlisting}
\end{appendix}
\end{document}